\documentclass[sigconf]{acmart}
\renewcommand\footnotetextcopyrightpermission[1]{} % removes footnote with conference information in first column
\def\BibTeX{{\rm B\kern-.05em{\sc i\kern-.025em b}\kern-.08emT\kern-.1667em\lower.7ex\hbox{E}\kern-.125emX}}
    
\usepackage{nicefrac}
\usepackage{siunitx}
\usepackage{array,framed}
\usepackage{booktabs}
\usepackage{
  color,
  float,
  epsfig,
  wrapfig,
  graphics,
  graphicx,
  subcaption
}
\usepackage{textcomp,amssymb}
\usepackage{setspace}
\usepackage{latexsym,fancyhdr,url}
\usepackage{enumerate}
\usepackage{algpseudocode}
\usepackage{graphics}
\usepackage{xparse} % argument parsing -- \edist
\usepackage{xspace}
\usepackage{multirow}
\usepackage{csvsimple}
\usepackage{balance}
\usepackage{
  tikz,
  pgfplots,
  pgfplotstable
}
\usepackage{hyperref}

\usetikzlibrary{
  shapes.geometric,
  arrows,
  external,
  pgfplots.groupplots,
  matrix
}

\pgfplotsset{compat=1.9}
\usepackage{mathtools,}

\DeclareMathAlphabet{\mathcal}{OMS}{cmsy}{m}{n}
\DeclareGraphicsExtensions{%
    .png,.PNG,%
    .pdf,.PDF,%
    .jpg,.mps,.jpeg,.jbig2,.jb2,.JPG,.JPEG,.JBIG2,.JB2}

\usepackage{xparse}
\newcommand{\bnm}{\begin{newmath}}
\newcommand{\enm}{\end{newmath}}

\newcommand{\bea}{\begin{eqnarray*}}%
\newcommand{\eea}{\end{eqnarray*}}%

\newcommand{\bne}{\begin{newequation}}
\newcommand{\ene}{\end{newequation}}

\newcommand{\bal}{\begin{newalign}}
\newcommand{\eal}{\end{newalign}}

\newenvironment{newalign}{\begin{align}%
\setlength{\abovedisplayskip}{4pt}%
\setlength{\belowdisplayskip}{4pt}%
\setlength{\abovedisplayshortskip}{6pt}%
\setlength{\belowdisplayshortskip}{6pt} }{\end{align}}

\newenvironment{newmath}{\begin{displaymath}%
\setlength{\abovedisplayskip}{4pt}%
\setlength{\belowdisplayskip}{4pt}%
\setlength{\abovedisplayshortskip}{6pt}%
\setlength{\belowdisplayshortskip}{6pt} }{\end{displaymath}}

\newenvironment{newequation}{\begin{equation}%
\setlength{\abovedisplayskip}{4pt}%
\setlength{\belowdisplayskip}{4pt}%
\setlength{\abovedisplayshortskip}{6pt}%
\setlength{\belowdisplayshortskip}{6pt} }{\end{equation}}

\newcounter{ctr}

\newcounter{mytable}
\def\mytable{\begin{centering}\refstepcounter{mytable}}
\def\endmytable{\end{centering}}

\newcounter{myfig}
\def\myfig{\begin{centering}\refstepcounter{myfig}}
\def\endmyfig{\end{centering}}

\newlength{\saveparindent}
\newlength{\saveparskip}
\newcommand{\E}{{\rm I\kern-.3em E}}

\renewcommand{\eqref}[1]{\mbox{Equation~(\ref{#1})}}

\def \part {part}

\renewcommand{\paragraph}[1]{\vspace*{6pt}\noindent\textbf{#1}\;}

\def \blackslug{\hbox{\hskip 1pt \vrule width 4pt height 8pt
    depth 1.5pt \hskip 1pt}}
\def \qed{\quad\blackslug\lower 8.5pt\null\par}
\newcounter{mynote}[section]

\newcommand\ignore[1]{}

\newcounter{rcnote}[section]

\newcounter{mrnote}[section]

\newcounter{fknote}[section]

\newcounter{anote}[section]

\DeclareMathSymbol{\mlq}{\mathord}{operators}{``}
\DeclareMathSymbol{\mrq}{\mathord}{operators}{`'}

\newcommand{\rhf}[2]{R_{f, \gamma}}

\DeclareDocumentCommand{\edist}{o o}{
  \ensuremath{
    \IfNoValueTF{#1}{{d}}{{\sf d}(#1,#2)}
  }
}

\newcommand{\olrk}[1]{\ifx\nursymbol#1\else\!\!\mskip4.5mu plus 0.5mu\left(\mskip0.5mu plus0.5mu #1\mskip1.5mu plus0.5mu \right)\fi}

\NewDocumentCommand{\indseq}{ O{1} O{r} }{{#1}\ldots {#2}}

\usepackage{algorithm}
\usepackage{xcolor}
\definecolor{azure}{rgb}{0.54, 0.17, 0.89}

\usepackage{enumitem,kantlipsum}

\newcommand{\tfp}{{Free2Shard } }
\newcommand{\tfpnosp}{Free2Shard\ignorespaces}

\usepackage{mathtools}
\usepackage{amssymb}
\usepackage{hyperref}

\begin{document}
%\fontfamily{lmr}\selectfont
% \def\thetitle{A Practical Way to Generate Strong Keys from Noisy Data}
\fancyhead{}
\def\thetitle{\tfpnosp: Adaptive-adversary-resistant sharding via Dynamic Self Allocation}
\title{\thetitle}

\author{Ranvir Rana}
\email{rbrana2@illinois.edu}
\affiliation{%
  \institution{University of Illinois at Urbana-Champaign}
}

\author{Sreeram Kannan}
\email{ksreeram@uw.edu}
\affiliation{%
  \institution{University of Washington at Seattle}
}

\author{David Tse}
\email{dntse@stanford.edu}
\affiliation{%
  \institution{Stanford University}
}

\author{Pramod Viswanath}
\email{pramodv@illinois.edu}
\affiliation{%
  \institution{University of Illinois at Urbana-Champaign}
}

\date{}

\begin{abstract}
  Propelled by the growth of large-scale blockchain deployments, much recent progress has been made in designing {\em sharding} protocols that achieve throughput scaling linearly in the number of nodes. However, existing protocols are not robust to an adversary {\em adaptively} corrupting a fixed fraction of nodes. In this paper we propose \tfpnosp -- a new architecture that achieves near-linear scaling while being secure against a fully adaptive adversary. 
  
  The focal point of this architecture is a dynamic self-allocation algorithm that lets users allocate themselves to shards in response to adversarial action, without requiring a central or cryptographic proof. This architecture has several attractive features unusual for sharding protocols, including: (a) the ability to handle the regime of large number of shards (relative to number of nodes); (b)  heterogeneous shard demands; (c) requiring only a small minority to follow the self-allocation;  (d) asynchronous shard rotation; (e) operation in a purely identity-free proof-of-work setting. The key technical contribution is a deep mathematical connection to the classical work of Blackwell in dynamic game theory. 
\end{abstract}

\maketitle

% Section I
\section{Introduction}
\label{sec:intro}

A classical problem in distributed systems is one of maintaining a state machine given $N$  nodes, some fraction of which are adversarial (also termed Byzantine). Classical  mechanisms for Byzantine-fault-tolerant (BFT) state-machine-replication (SMR) rely on full replication of data across multiple nodes; thus offering no scaling in efficiency as the number of replica nodes increase \cite{castro1999practical,dolev1983authenticated,dwork1988consensus,abraham2019communication}. Since SMR is the key primitive underlying blockchains, it is no surprise that the first generation of blockchains also relied on full replication \cite{bitcoin, eth2}. 

Given the unprecedented scale of blockchains, for example, Bitcoin has $N \geq 10,000$ nodes running its protocol \cite{btc10K}, there is compelling practical interest in protocols whose efficiency increases with the number of nodes. This problem has attracted wide interest in the distributed systems community with many protocols being proposed \cite{luu2016secure,kokoris2018omniledger,zamani2018rapidchain}. These pioneering methods offer provable security as well as near-linear scaling  in $N$ of efficiency across various resources at an individual node including computation, storage and communication.

While existing solutions offer excellent performance as well as security against static (or slowly adapting) adversaries, their security fails under an adaptive adversary. This problem is compounded in permissionless blockchain deployments where the possibility of the adaptive adversary threat is high since nodes do not have persistent identities. This is evidenced in the significant practical interest in designing protocols robust to this setting \cite{zilliqa,wood2016polkadot,eth2,lazyledger}.  

The main result of this paper is the \tfp architecture, whose performance scales near-linearly with the number of nodes while being secure against {\em fully adaptive adversaries} controlling up to 50\% of the nodes. While existing sharding solutions build upon a cryptographically certifiable node-to-shard allocation algorithm, we take the complete opposite view: nodes can allocate themselves to shards as they please. Our sharding architecture is designed in such a way that even when the majority of a shard is adversarial, the safety is not violated. However, an adversary can congregate in a shard, significantly reducing the fraction of honest shard blocks, and creating corresponding security threats (especially, liveness) and also  restricting throughput. This is solved by our core contribution: a {\em dynamic shard allocation algorithm}. 

The core idea underneath our algorithm is that the honest nodes  re-allocate themselves into shards throttled by the adversary. However, the adversary can observe the honest nodes's actions and re-allocate itself to nullify the honest nodes' actions. The main technical contribution of this paper is the identification of a (computationally simple) dynamic self-allocation policy that can successfully ensure that the fraction of honest to adversarial nodes in {\em every shard} is greater than $0.5$ (thus ensuring sufficient throughput in all shards). The core technical result is a complete and striking solution to a dynamic Stackelberg game \cite{stackelberg}; our approach is distinct but inspired by  the classical Blackwell approachability in game theory \cite{blackwell1956analog}.  

The \tfp architecture also uses well-established primitives to achieve its requisite properties: (1) an SMR engine, resistant to adaptive adversaries, to maintain an ordered log of all shard block hashes, (2) a data availability engine \cite{al2018fraud,CMT} that guarantees that data written into the SMR are actually available, and a (3) state-commitment engine based on interactive verification \cite{truebit,kalodner2018arbitrum} to ensure low complexity of bootstrapping while rotating.  

Furthermore, the protocol has several desirable properties that makes it attractive from a systems view: (a) asynchronous shard rotation - nodes do not all rotate at the same time; (b) requires only a small number of honest nodes per shard; (c) guarantees the aforementioned properties even when only a small minority of nodes follow the proposed rotation protocol; (d) can support heterogeneous shard throughput even when the number of shards is greater than the number of nodes;  (e) the ability to operate in a (permissionless) proof-of-work setting with the same guarantees of throughput scaling and security against a fully adaptive adversary. 

The paper is organized as follows. There is a very large number of recent literature on sharding solutions for blockchains and we conceptually organize them in terms of their architecture (and building blocks) in Section~\ref{sec:relwork}; this survey also sets the stage to put the proposed \tfp architecture in context (and provide a brief overview).  The key building block of \tfp is the dynamic self allocation engine which is discussed in detail in Section~\ref{sec:dsa}. We describe \tfp architecture and various parameter choices in Section~\ref{sec:tfp}; we also provide the security guarantees and efficiency scaling properties. We conclude in Section~\ref{sec:systemView} with a discussion on the  various system properties of \tfp in the context of real world implementation concerns (especially in a  distributed permissionless setting) --  synchronization, partial deployment, incentives, inter-shard transaction handling.

\section{Background and Motivation}
\label{sec:relwork}

\noindent\textbf{Security Model}. 
We consider a distributed system maintained by $N$ nodes on a synchronous network.  We assume that there is a natural mechanism to partition the ledger into $K$ distinct shards of equal size (for example, think of these as distinct applications sharing a common blockchain or as distinct accounts in a payment system). The goal is to have the total throughput {\em scaling linearly with the number of nodes}, while each node expends only {\em constant amount of resources}. We will particularly be concerned about three types of resources: (1) {\bf computation} resource - the number of transactions executed per second, (2) {\bf storage}  resource - the number of transactions / second that can be incrementally stored by the node \footnote{We will make the simplifying assumption that the storage per node is growing with time. This is required for Bitcoin for example. Ideas to relax this requirement, for example \cite{jedusor2016mimblewimble}, can be naturally applied to our setting too.} and (3) {\bf communication}  resource - the total amount of data communicated by a node. We will assume that at every node, \textit{ all three resources are sufficient} to run a non-sharded blockchain with $R$ transactions-per-second. We will assume that, at any given time, a fraction $\beta$ of nodes are controlled  by the adversary - they can deviate arbitrarily from the proposed protocol. We want to prove security against a {\em fully adaptive adversary}, which can corrupt any subset of nodes based on the public state till that time (as long as the total number of corrupted nodes is less than its ``budget" $\beta N$). Security encompasses two aspects: (1) {\em safety}: transactions once confirmed remain confirmed for ever, and (2) {\em liveness}: new honest transactions will continue to be added within a finite amount of time. 

This problem has elicited much recent interest owing to the scalability bottleneck in blockchains; the corresponding solutions in the literature   are broadly referred to as sharding methods. Blockchain sharding methods in the literature  can be broadly divided into two categories: multiconsensus (most sharding solutions) and uniconsensus architectures. Each of these architectures provide different scaling gains and have distinct security vulnerabilities; we discuss this next. 

\subsection{Multiconsensus architecture}

\begin{figure}
    \centering
    \includegraphics[width = .48\textwidth]{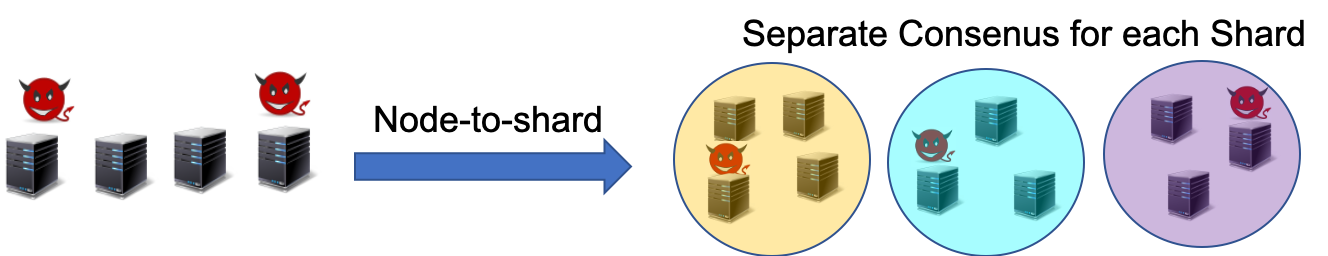}
    \caption{Multiconsensus architecture}
    \label{fig:n2s1}
\end{figure}

The multiconsensus architecture relies on each shard having a {\em separate consensus engine} and  has a secure cryptographic allocation of nodes randomly to shards. A further periodic random reallocation of nodes to shards protects against a weakly adaptive adversary. The random reallocation is enabled  by a {\em node to shard allocation engine} (N2S) which assigns nodes to shards using a (common) distributed randomness, generated using a  consensus engine shared across all shards and commonly referred to as  {\em beacon consensus engine}. A {\em state commitment engine} posts root of a Merkle Trie \cite{luu2017peacerelay} of a shard's execution state (the so-called ``state commitment") \cite{eth2} to the beacon consensus engine at regular intervals; this facilitates fast reallocation  and  synchronization to new shards. Figure \ref{fig:n2s1} illustrates a simple example with a node-to-shard allocation that allocates nodes to distinct shard, on which independent consensus is performed.

 {\em Security}. If a (super)majority exists in the overall set of nodes, then the random node to shard allocation engine transfers this property to each shard (as long as each shard is large enough -- this weakness is discussed in detail shortly). This guarantees security of the shard consensus engine, against a static adversary. Periodic reallocation of nodes to shards enhances the security of shard consensus against a slowly adaptive adversary.  

{\em Scaling}. 
Multiconsensus architecture allows nodes to maintain only the state of the shard consensus engine and the beacon consensus engine, with beacon consensus engine only containing state commitments and N2S allocation metadata. Splitting the set of nodes into $K$ different shards, enables parallel execution of $K$ shard consensus engines, thus scaling transaction execution by factor $K$.

A key question is how large can $K$ be, relative to $N$. With a random allocation of node to shards, a majority fraction of honest nodes overall (eg: 60\%) translates to majority of honest nodes in a given shard with high probability (eg. $10^{-10}$) only if the size of the shard is large (i.e., $G=1100$ nodes). Since each shard should have $G$ members, the total number of shards $K$ has to be smaller than $N/G$ (i.e,, $N/1100$). This restricts the number of shards and thus the scaling capability.

Elastico \cite{luu2016secure} and Omniledger \cite{kokoris2018omniledger} pioneered the study and design of sharding methodologies for blockchain in academic literature. Omniledger, built on top of ideas from Elastico, runs N2S allocation using Randhound(a randomization protocol) on an identity blockchain. The nodes are gradually rotated between shards to decrease the synchronization load on the network. Rapidchain \cite{zamani2018rapidchain} further proposed several improvements by reducing the communication complexity to sub-linear in the number of nodes, and secure reallocation of nodes to shards building on the Cuckoo rule.

{\em Key Vulnerability}. The key property enabling security of the multiconsensus architecture is that there is sufficient honest (super)majority in {\em each} shard (this suffices for a static adversary) and the N2S allocation is periodically updated (this allows security against a weakly adaptive adversary). However the scheme is insecure against adaptive adversaries:
 The N2S allocation is posted on the beacon chain, hence an adversary  knows the list of nodes participating in any given shard. An adversary can then adaptively target all nodes allocated to a  particular shard and completely corrupt it (this is well within the corruption budget, since the total number of nodes in a shard is small relative to the network size). Once a shard is taken over by the adversary, both safety (i.e., the adversary can approve invalid blocks) and liveness (the adversary can block honest transactions) can be compromised. 
 
 In permissionless settings, the possibility of the adaptive adversary threat is severe. This is reflected in the fact that permissionless sharding protocols have designed heuristic mechanisms to deal with this threat. Consider the case of Ethereum 2.0 \cite{eth2} or Polkadot \cite{wood2016polkadot}. In these protocols,  the primary mechanism for dealing with adaptive adversaries is to submit commitments of shard state into the beacon consensus engine, and any node can contest this commitment by proving that the state transition from previous commitment includes an invalid transaction. This short proof is called as a fraud proof and can be posted by anyone to invalidate a set of shard blocks \cite{al2018fraud}. 
 
 While fraud proofs can be used to detect safety violations caused due to an adaptive adversary, they cannot detect liveness violations.  Indeed, an adaptive adversary can corrupt the majority of any given shard and ask them to censor all honest transactions. Thus since no invalid transactions have been included in the ledger, no fraud-proof can be created. Every time the node-to-shard allocation is rotated, the adaptive adversary corrupts the newly allocated members, thus imposing a permanent liveness ban on that shard. 

While cryptographic scaling alternatives have been proposed, we note that this liveness attack persists. For example, mechansims of verifiable computing \cite{parno2013pinocchio} such as SNARKs \cite{ben2014succinct}, ZK-STARKs \cite{ben2018scalable} and bullet-proofs\cite{bunz2018bulletproofs} have been proposed for scaling blockchains under differing assumptions on trust and setup. These mechanisms enable scaling by letting the block proposer guarantee that the posted state accurately reflect the state after executing the transactions in the block. While these mechansims can be used as a non-interactive alternative to fraud-proofs for scaling, they have no way of guaranteeing that transactions have been censored, and are thus subject to the liveness attack by the adaptive adversary described above.

\subsection{Uniconsensus architecture}
\label{sec:uniconsensus}
\begin{figure}
    \centering
    \includegraphics[width = .4\textwidth]{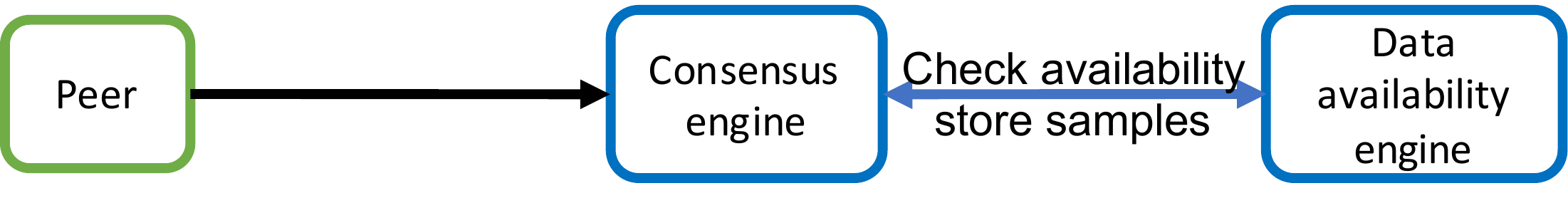}
    \caption{Uniconsensus architecture}
    \label{fig:trivial_sa}
\end{figure}

The requirement for node-to-shard allocation in multiconsensus architecture comes from the necessity of allocating nodes to each of the shard consensus groups. The uniconsensus architecture avoids this requirement by instead relying on a single {\em consensus engine} to maintain the ordering information of {\em all} shard blocks. Thus the safety of each shard can be directly derived from the safety of the main consensus engine. The consensus engine only maintains a {\em log of hash of shard blocks} and hence is scalable. A honest consensus (super)majority is no longer needed in each shard to preserve safety, thus each node is free to join a shard of it's choice (self-allocation). Shard block execution is performed by shard nodes and is decoupled from shard block ordering performed by the consensus engine. This architecture does not require each transaction in every shard block to be valid, rather only guaranteeing the order of transactions in the shards. Since the consensus engine only maintains the hash of each block, a separate mechanism is required to ensure that the full block is available to the shard nodes - this is guaranteed by a {\em data availability engine}.  Figure \ref{fig:trivial_sa} illustrates the 3 engines and their interactions.

{\em Security}.  If a  majority of honest nodes exist in the overall set of nodes, the consensus engine is secure. Consider a shard with less than a majority of honest nodes, the adversarial nodes in the shard cannot change the log of shard blocks without violating safety of the consensus engine, thus an adaptive adversary with less than a global consensus majority cannot violate safety of any shard. The lack of requirement of a honest majority constraint within a shard also allows for small shard size. We observe that uniconsensus architecture solves the shard ledger safety and shard size vulnerabilities of multiconsensus architecture, but introduces serious liveness vulnerabilities discussed below. 

{\em Scaling}.  Nodes maintaining the uniconsensus architecture only maintain the consensus engine and shard log of one shard, the consensus engine contains a log of shard block hashes, the size of which is much smaller than shard blocks. Thus, $K$ shards can be processed in parallel with each node processing only one shard log. 

{\em Related work}. 
The idea of scaling by a distributed system by maintaining a single consistent log with distributed data is propounded by distributed systems architectures like Corfu \cite{balakrishnan2012corfu} and Tango \cite{balakrishnan2013tango}. Corfu proposed an architecture to maintain a log whose data is distributed across a cluster of flash drives. Tango built a sharded system on top of Corfu, where execution is decoupled from validity and application nodes only need to execute the subset of entries from the log which are relevant to that application. At a high level, this fits into the uniconsensus architecture described here, but Corfu and Tango are designed for resilience to the simpler crash-faults as compared to the more complex Byzantine faults considered in this paper.

To handle more complex fault models, works like Aspen \cite{gencer2017short} require consensus nodes to maintain and compute the entire ledger data. However, the data is organized like in Tango, so that client nodes running an application only need to download the relevant data. Lazyledger  \cite{lazyledger} took this idea further to reduce storage and communication burden on the consensus nodes by allowing data availability proofs \cite{al2018fraud}.

\begin{figure}
    \centering
    \includegraphics[width = .4\textwidth]{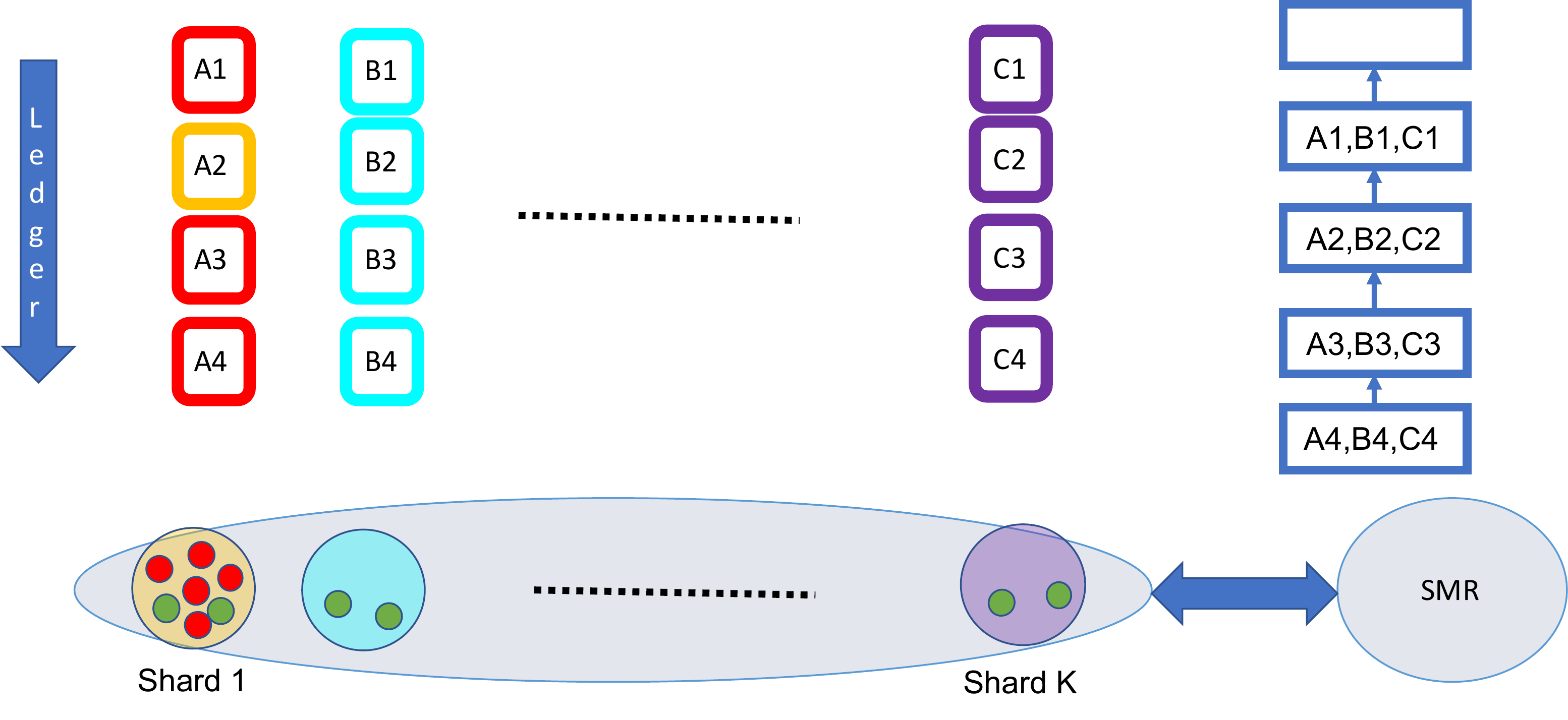}
    \caption{Throughput suppression: Adversaries congregate in shard 1. }
    \label{fig:tp_suppression}
\end{figure}

{\em Vulnerability}.  Removing N2S allocation by allowing node self-allocation introduces a serious liveness attack on a shard where an adversary can concentrate it's mining power on one shard and drown out honest shard blocks. We note that to execute this liveness attack, we do not even require an adaptive adversary. In this attack, the fraction of honest blocks to the adversarial blocks is greatly reduced due to a large fraction of adversarial nodes in the shard. Since the shard has limited resources, it can only support a fixed number of shard blocks, hence a low honest block to total block fraction would lead to a throttling of the number of honest shard blocks added to the shard log and hence throttle throughput. The honest transactions can no longer be processed in finite time, thus leading to loss of liveness. An example of this attack is shown in Figure \ref{fig:tp_suppression}.

\subsection{Partial Scaling Approaches} The approaches considered till now achieve scaling while maintaining constant resource usage, i.e., these protocols are computation, communication and storage efficient. There are other approaches that are efficient only in some dimensions while being fully resilient to an adaptive adversary. For example, Zilliqa \cite{zilliqa} achieves computation-efficient scaling, but it is not communication or storage efficient. Polyshard \cite{li2018polyshard} and coded state machines \cite{li2019coded} can achieve storage efficient scaling but they are not communication efficient. 

To summarize: (1) the multiconsensus sharding architecture is safe against weakly adaptive adversaries, but loses safety and liveness against adaptive adversaries,  (2) the uniconsensus sharding architecture is safe against adaptive adversaries but liveness is compromised even under static adversaries, and (3) approaches fully secure against an adaptive adversary are not efficient in all dimensions.

\subsection{Overview of \tfp}
\label{sec:overview}

The main contribution of this paper is the introduction of the \tfp architecture which provides full horizontal scaling (in all the 3 dimensions of storage, compute and communication) while being secure against fully adaptive adversaries. The architecture builds on the uniconsensus architecture described above with a new component: the {\em dynamic self-allocation (DSA) engine}.  The DSA engine provides an algorithm for honest nodes to re-allocate themselves to shards, thus reacting to adversaries congregating in individual shards, and guarantees a strong mathematical property: the {\em time average} fraction of honest to adversarial nodes in every shard approaches the theoretical optimal (fraction of all honest nodes, total across all shards). Deriving this dynamic self-allocation engine and its mathematical guarantee is a core algorithmic and theoretical contribution of this paper and is discussed in detail in the next section.

\begin{figure}
    \centering
    \includegraphics[width = .48\textwidth]{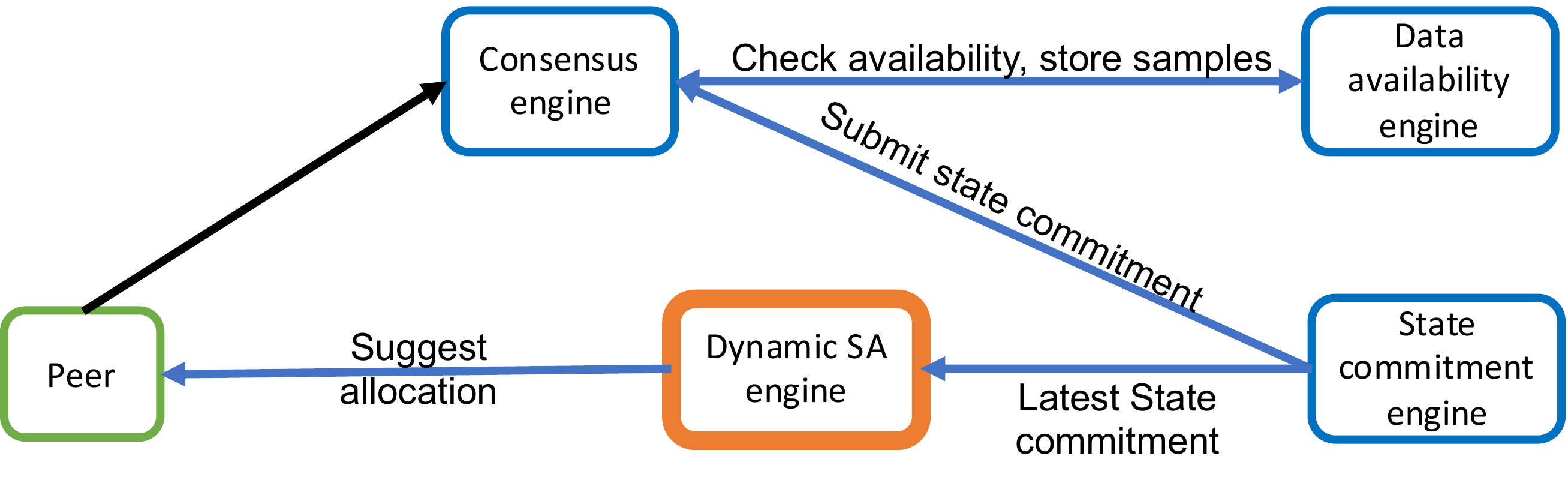}
    \caption{Free2shard Architecture}
    \label{fig:overview}
\end{figure}

\begin{figure}
    \centering
    \includegraphics[width = .48\textwidth]{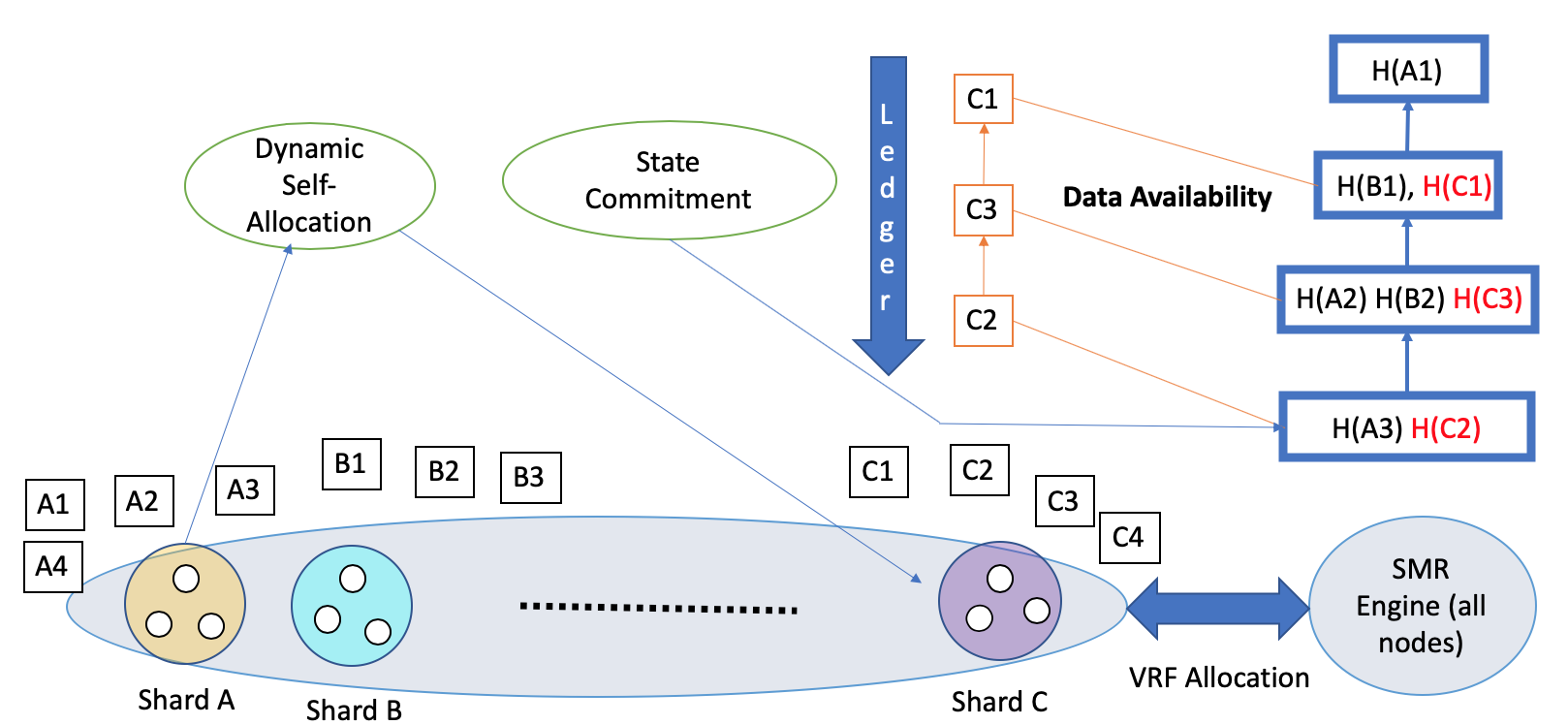}
    \caption{Free2shard System in Action}
    \label{fig:overview2}
\end{figure}

The \tfp architecture comprises of the following: (1) State machine replication (SMR) consensus engine run by all nodes for maintaining a total order of block hashes, (2) data availability engine to certify that every block hash written into the SMR is actually available in that shard, (3) dynamic self-allocation (DSA) engine to ensure that the fraction of honest nodes in any given shard is high and (4) state commitment  engine, that provides nodes migrating between shards a quick boostrapping ability to get the state of the new shard without requiring them to download the entire other shard.  We note that the primitives (1), (2) and (4) are independently well-understood from the literature and have been used in different ways in sharding architectures. The unique properties of \tfp arise primarily from the DSA and the way in which it leverages the other primitives. Figure \ref{fig:overview} illustrates the 5 engines and their interactions and Figure \ref{fig:overview2} is an example of a Free2Shard system in action. At a high level, the DSA remedies the liveness vulnerability of the uni-consensus architecture. 

\subsection{Key Primitives}
\label{sec:keyprimitives}
We require three key primitives from previous literature in our \tfp architecture, we describe these primitives briefly here.

 {\bf State Machine Replication}. A totally ordered log of the shard block headers is maintained using a SMR \cite{bessani2014state}. We assume that the SMR is robust to an adversary adaptively corrupting at most $\beta_{c}$ fraction of the nodes. To achieve scalability, we require that the total message complexity of SMR be linear in the number of nodes (so that the communication complexity per node is a constant).  
 
 Examples of SMR that satisfy these properties include Ouroboros Praos and Algorand, since they are robust to fully adaptive adversaries. We note that Ouroboros Praos has $\beta_c = 0.5$ whereas Algorand has $\beta_c=\frac{1}{3}$. Other protocols which have efficient communication complexity such as HotStuff \cite{yin2019hotstuff}, Sync HotStuff \cite{abraham2019sync}, and Streamlet \cite{chan2020streamlet} are robust to varying degrees of adaptivity by the adversary. 

 {\bf Data Availability engine}. We require data availability engine as a key primitive.  \textit{Input}: Commitment $c$ for a shard block (of size $B$) from a honest / adversarial node. \textit{Output}: Every node deduces correctly if the block is available. 
 \textit{Properties}:  Nodes inside that shard expend at most $O(B \log B)$ resources. Every node expends at most $O(\log B)$ resources.
  \textit{Assumption}: Correct if at least one honest node inside the shard.

\textbf{State commitment engine}. The state commitment engine is used to commit the execution state of a shard into the SMR after executing a block of transactions.  We use the interactive state verification engine from \cite{truebit} that satisfies the following properties. We note that verifiable computing methods such as ZK-STARK \cite{ben2018scalable} can be used instead too.  
\textit{Input}: Prior state commitment $c$, block $B$ of transactions, claimed state $s$ after executing $B$ beyond $c$. The claimed state can be posted either by a honest node or an adversary. \textit{Output}: Every node correctly agrees on whether $s$ is correct or not. \textit{Properties}: Requires a distinct committee for $O(\log B)$ rounds. Every committee member spends at most $O(N \log B)$ resources. All other nodes expend at most $O(\log B)$ resources. 
 \textit{Assumption}: Correct if there is at least one honest node in the committee of each round.

%Shard log is derived from the consensus engine and maintained by shard nodes. The shard nodes change their allocation dynamically by calling the DSA engine locally at regular time intervals. The nodes synchronize to the newly allocated shard using the state commitments from the state commitment engine which works in a similar manner as its counterpart in multiconsensus architecture. 

\subsection{Our Contribution}
{\bf Key Properties} \tfp satisfies the following properties: 
\begin{itemize}
    \item \tfp is safe as long as $\beta$ (fraction of nodes are controlled by the adversary) is less than  $\beta_c$, the critical threshold of fraction adversaries that can be tolerated by the underlying SMR consensus engine. 
    \item \tfp is live as long as $\gamma > 0$ fraction of nodes follow the self-allocation policy. 
    \item \tfp can support $K < \Theta(\frac{N} {\log N})$ shards each of throughput at least $R/2$ transactions per second. Thus we get a total scaling of $\Theta(\frac{N} {\log N})$ relative to running a single shard.
\end{itemize}

We provide a short summary of why \tfp achieves the security and scalability properties below. A detailed discussion is deferred to Section~\ref{sec:tfp}.

\begin{itemize}
    \item Each shard block is safe as long as the SMR engine is safe, which is true till $\beta<\beta_c$ for the SMR even under an adaptive adversary. 
    \item The shard continues to include honest transactions as long as honest nodes get to propose blocks. Since our DSA guarantees that the long term fraction of honest blocks in a given shard is close to $0.5$, liveness is guaranteed even against an adaptive adversary. Furthermore, we show that even a constant fraction $\gamma$ of nodes following the DSA is sufficient to ensure a long term fraction of honest blocks in any shard is close to $\gamma$. 
    \item The overhead for each node associated with a block of size $B$ in any shard is $O(\log B)$ since the overhead of each primitive is $O(\log B)$. SMR requires $O(\log B)$ overhead to read the Merkle proof of each chunk, data availability and state commitment also require $O(\log B)$ resources as stated in the primtives above. The total overhead ratio is $O(\frac{K \log B}{B})$, which can be made arbitrarily small by making $B$ large.
\end{itemize}

%{We require the remaining $1-\beta$ honest nodes to follow the underlying SMR consensus protocol but they need not comply with the self-allocation policy.}

\section{Dynamic Self Allocation Engine}
\label{sec:dsa}

The dynamic self allocation engine is a sequential algorithm that guides honest nodes to allocate themselves to shards in different proportions, adapting to the past adversarial allocation behavior. The DSA algorithm aims to guarantee that the time average fraction of honest to adversarial nodes in any shard is above any desired level. A fundamental information theoretic question is a characterization of the space of time average fractions that can be achieved by the best DSA algorithm (here we do not constrain the computational complexity of algorithm). A practical, and also theoretical, question is the characterization of the performance of computationally simple DSA algorithms and the explicit identification of such algorithms. We formally state these questions below; the main result of this section is a complete solution to these theoretical questions -- this comprises  the key technical contribution of this paper. We discuss connections of our  results to game theoretic literature (dynamic Stakelberg games \cite{stackelberg}) and is of independent mathematical interest.

%Our protocol allows peers to join any shard they want without any central consensus engine guiding them. We discussed in section \ref{sec:relwork} that following a trivial self allocation policy leads to a liveness attack. We then propose a few policies that solve the problem, with our final policy ensuring that independent non-coordinating honest peers can completely nullify the effects of a coordinating adversary within a constant factor $h$. We show that our final DSA policy works even for $N<K$.

{\bf Problem Statement}. There are $N$ nodes (including both honest and adversarial) and $K$ shards. The fraction of honest nodes is $(1-\beta)=\gamma$ (the ``power" of honest nodes) and each honest node uses the DSA algorithm to allocate itself to a shard. The adversaries collude, observe the honest node allocations and then allocate themselves. The goal of the adversary is to minimize the ratio of honest to total nodes (the ``honest fraction") in any shard -- this will throttle the throughput of the shard leading to liveness vulnerabilities (as discussed in Section~\ref{sec:uniconsensus}).  The honest nodes switch at periodic intervals (time scale of confirmed blocks in the consensus engine). Even though the adversarial nodes can switch faster, we see that they cannot reduce the time average honest fraction of the worst performing shard further than reallocating themselves at the same rate as honest nodes. This leads to the following mathematical formulation of a vector dynamic game described below, focusing on the {\em mean} honest fraction allocations to each shard. The actual DSA algorithm is to be implemented by every node and is necessarily distributed -- a randomized distributed implementation (with the goal of mimicking the mean honest fraction prescribed by the solution to the dynamic game below) is discussed in a later Section~\ref{sec:ddsa}.  %\ref{eqn:main_function}

We suppose a timescale where the consensus engine confirms one block per unit time. We measure the honest node fractions in shard at integer times $t$; this fraction  would correspond to the ratio of honest blocks to total blocks in any shard, formed in time $[t-1,t]$ referred by the consensus block confirmed at time $t$. We refer to the interval $[t-1,t]$ as round $t$. Let $r_i[t]$ denote the honest node fraction at time $t$ (for round $t$) in shard $i$ and $\bar{r}_i(t)$ denote the time-average.
\begin{align}
    r_i(t) &= \frac{\gamma_i(t)}{\gamma_i(t) + \beta_i(t)}; \quad
    \bar{r}(t) = \frac{1}{t}\sum_{j=1}^t r_i(j).
\end{align}
Let $\mathbf{r}(t) = [r_1(t), r_2(t), .., r_K(t)]$ and $\mathbf{\bar{r}}(t) = [\bar{r}_1(t), \bar{r}_2(t), .., \bar{r}_K(t)]$ denote the corresponding vectors encompassing all the $K$ shards.

We define the main optimization objective function, the solution of which is the honest strategy that can achieve a worst case shard honest fraction of $\psi(K)$ given any worst-case adversarial action:
\begin{equation}
   \psi(T) = \max_{\{f_t\}_t} \, \min_{\{\beta_i(t)\}_{i,t}} \, \min_i \left\{  \bar{r}_i(T) \right\}. \label{eqn:main_function}
\end{equation}

A DSA algorithm specifies the honest node re-allocation strategy: for each $t$ the re-allocation is a function  as follows:
\begin{equation}
  f_t: (\boldsymbol{\beta}(1),...\boldsymbol{\beta}(t-1),\boldsymbol{\gamma}(1),\ldots\boldsymbol{\gamma}(t-1)) \rightarrow \boldsymbol{\gamma}(t),
\end{equation}
satisfying the constraint $\sum_{i=1}^K \gamma_i(t) = \gamma$.

We define $\psi_{f_t}(K)$: the honest fraction of the worst performing shard under a worst case adversary as follows:
\begin{equation}
   \psi_{f_t}(T) = \, \min_{\{\beta_i(t)\}_{i,t}}    \, \min_i \left\{ \bar{r}_i(T) \right\}. 
\end{equation}
Note that the adversary action in round $t$ is allowed to depend on the  honest node re-allocation policy at round $t$.

{\bf Information Theoretic Limits}. 
We see  that any optimal honest strategy cannot obtain a time averaged honest fraction greater than $\gamma$ in the worst performing shard under an optimal adversarial strategy: 
\begin{align}
    \psi(T) \leq \gamma. \label{eq:info_theoretic_1} %\max_{\{f_t\}_t} \, \min_i \left\{ \frac{1}{T} \sum_{t=1}^T \frac{\gamma_i(t)}{\gamma_i(t)+\beta_i(t)} \right\}. 
\end{align}
The above inequality is obtained since irrespective of the honest allocation, an adversary can simply replicate the honest node allocation and set $\beta_i(t)=\frac{\beta}{\gamma}\gamma_i(t)$, thus rendering the honest fraction of each shard as $\gamma$.
%The right hand side in the above equation fixes an adversarial strategy $\{\beta_i(t)\}_{i,t}$. The optimal value in the right hand side is $\gamma$ and is achieved when $\gamma_i(t)=\frac{\gamma}{\beta}\beta_i(t)$. Please refer to the appendix for detailed proof.
The key question is whether this upper bound is actually achievable.  If so, then this would be a striking and positive result: the honest nodes can rebalance the allocations optimally, i.e.,  the time average fraction of honest nodes in each shard is always the same as the overall fraction of honest nodes (which is $\gamma$). This would imply that the adversary is ineffective in overcrowding any shard, even minimally. The main result of this section is that this is indeed achievable, for large enough time window $T$ over which the average is taken.   Towards building an intuition towards this result, and the  nature of the optimal honest policy and the corresponding performance, we discuss next  a few baseline honest node strategies and analyze their performance.

\subsection{Approach} 

{\bf Static strategies}. 
Consider a static self-allocation scheme where honest nodes allocate themselves uniformly to one of $K$ shards, i.e., $\gamma_i(t) = \frac{\gamma}{K} \forall i \in [K]$. This strategy is very vulnerable to adversarial action:  an adversary simply targets shard $1$ and throttles its honest fraction by allocating all its  power to this shard (i.e.,  $\beta_1(t)=\beta$). This renders the honest fraction of shard 1 as $\frac{\gamma/K}{\gamma/K + \beta} = O(1/K)$ which approaches 0 for large $K$ and is very sub-optimal compared to the information theoretic limit. Thus the honest nodes have to adapt to adversarial action. 

\noindent {\bf A simple dynamic strategy}. 
A simple adaptive action by the honest nodes is the following: half the honest nodes distribute themselves uniformly randomly across $K$ shards and the other half follows the adversarial distribution of the past round.
\begin{align}
    f_{t} := {\gamma_i}(t) =    \{ \frac{\gamma}{2\beta} \beta_i(t-1) + \frac{\gamma}{2K} \}. 
    \label{eq:simpledynamic}
\end{align}
The idea is that if the adversary congregates in one shard, then at least half the honest nodes react to this congregation and ameliorate any degradation in performance in that shard. We are able to show the following performance result; the  proof is in Appendix~\ref{sec:proof_dsaa}.  
\begin{proposition} For large enough $T$, 
\begin{align}
    \Omega\left(\frac{1}{\log K}\right) \leq \psi_{f_{t}}(T)  \leq O \left(\frac{\log \log K}{\log K} \right).
    \label{thm:dsa1}
\end{align}
\end{proposition}
This shows that this dynamic allocation policy is much improved compared to the static allocation policy (from $O(1/K)$ to $O(1/\log K)$), but the performance still degrades to zero as $K$ grows, while the information theoretic limit is a constant (independent of $K$).  An adversarial attack on shard 1 which renders a honest fraction of $O(\frac{\log(K)}{\log(\log(K))})$ is illustrated in Figure \ref{fig:dsa1}. We see from the proof in Appendix~\ref{sec:proof_dsaa} that the analysis is very specific to the structure of the dynamic honest policy in  \eqref{eq:simpledynamic} and this does not provide any intuition towards designing an improved honest policy. We make connections to game theoretic literature next, which provides us a broader view of this problem and classical results associated with this class of $\max\min$ games. 

\begin{figure}
    \centering
    \includegraphics[width = .35\textwidth]{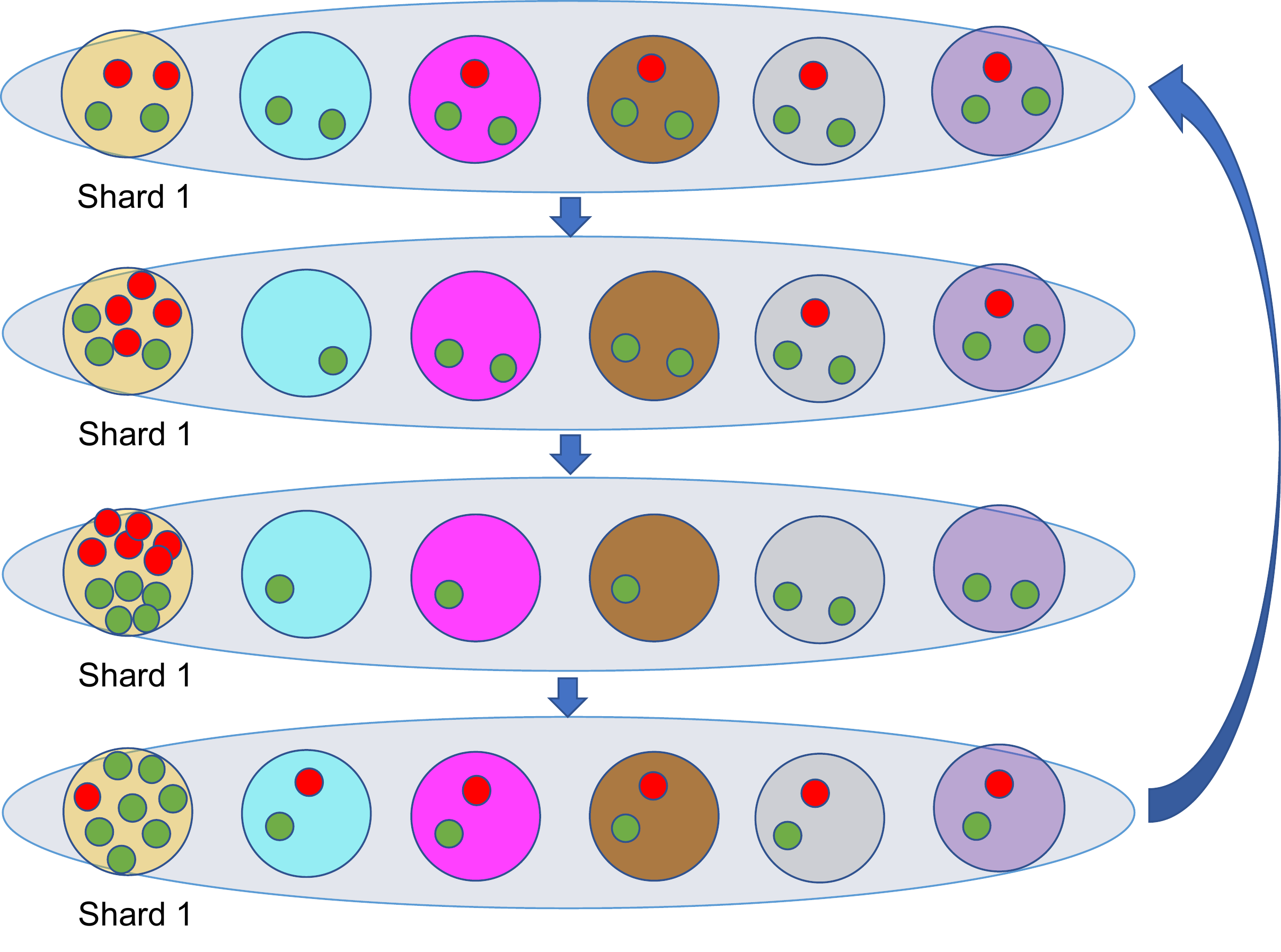}
    \caption{Simple DSA policy from  \eqref{eq:simpledynamic}: adversary gradually attacks shard 1, and honest nodes follow it with a delay of one round, the attack is reset when the adversary has allocated all it's  power to shard 1 }
    \label{fig:dsa1}
\end{figure}

%\paragraph{Lift-minimum strategy:} \textit{Honest strategy:} Focus half of all honest mining power to the shard with $\min_{i} \bar{r}_i(t)$ and the rest half uniformly amongst all shards, if that shard is shard $i$ $\gamma_i=\gamma/2 + \gamma/2K$. \textit{Claim:} $\psi_{f_{t}}(K)  \leq O(1/\sqrt{K})$. \textit{Proof:} Consider an adversarial strategy which focuses it's mining power in the lowest $\sqrt{K}$ shards, i.e. for those shards, $\beta_i(t)= \beta/\sqrt{K}$. Thus the throughput of the previous worst performing shard is given by $(\gamma/2 + \gamma/2K)/(\gamma/2 + \gamma/2K+\beta/\sqrt{K})$, however throughput of the rest $\sqrt{K}-1$ minimum shards is given by $(\gamma/2K)/(\gamma/2K+\beta/\sqrt{K}) = O(1/\sqrt{K})$. The adversary thus ensures that the shard with min throughput is amongst those $\sqrt{K}$ shards and the minimum of them only grows by $1/\sqrt{K}$.

\noindent {\bf Dynamic games}. 
 Our core optimization problem in \eqref{eqn:main_function} looks somewhat similar to an online convex optimization problem \cite{arora2012multiplicative}. However, the  combinatorial optimization $\min_i \bar{r}_i(T)$ outside the summation is a key aspect of departure from the online convex optimization setting, rendering  that approach nowhere immediate. An alternative view is provided by dynamic Stackelberg games \cite{stackelberg}, with zero step rewards and a terminal reward of $\min_i \bar{r}_i(T)$, where the adversary can take action after knowing the action of the honest nodes. However, solving the optimal honest policy iteratively (dynamic programming approach) does not yield an analytical form   due to the specific reward design we have. If the reward of $\min_i \bar{r}_i(T)$ were replaced by the vector  $\boldsymbol{\bar{r}}$, then this game is related to the classical Blackwell approachability \cite{blackwell1956analog} (an extension of von Neumann's classical minimax matrix game \cite{von1953certain}) and provides guidance as to how the scalar reward of $\min_i \bar{r}_i(T)$  be addressed. This is the crux of our main result, presented next. 
 
\subsection{Dynamic Self allocation} 
Consider the following \tfp policy, $f_t^{\textit{F2S}}$ where  $\gamma_i(t)$ is generated as follows:
\begin{eqnarray}
    f_t^{\textit{F2S}}: \gamma_i(t) = \gamma \frac{u_i(t-1)}{\sum_{i=1}^{K}u_i(t-1)}; \quad  \quad u_i(t-1) = (\gamma - \bar{r}_i(t-1))^+.  \label{eqn:main_policy} %\\ 
    %\text{where }\: u_i(t-1) = (\gamma - \bar{r}_i(t-1))^+. \nonumber
\end{eqnarray}

The main idea is to allocate honest  power to shards that are performing the worst, i.e., shards with highest lag from the target honest fraction $\gamma$ and not waste any honest  power by allocating  zero  power to shards which have time averaged  honest fraction greater than the target.   
Our main result is in the following theorem. 

\begin{theorem}
    For any  adversarial strategy, 
    \begin{align}
        \psi_{f_t^{\textit{F2S}}}(T) \geq \gamma\left(1 - \sqrt{\frac{K}{T}}\right).
    \end{align}
    \label{thm:tfp}
\end{theorem}
We observe that $\psi_{f_t^{\textit{F2S}}}(T) \geq c\gamma$ for all  
$T>\frac{K}{(1-c)^2\gamma^2}$, this means that we achieve $\psi_{f_t^{\textit{F2S}}}(T)$ is arbitrarily close to the information theoretic limit of $\gamma$ for $T$ large enough, thus providing tight bounds on approaching the information theoretic limit. 

\begin{figure}
    \centering
    \includegraphics[width = .3\textwidth]{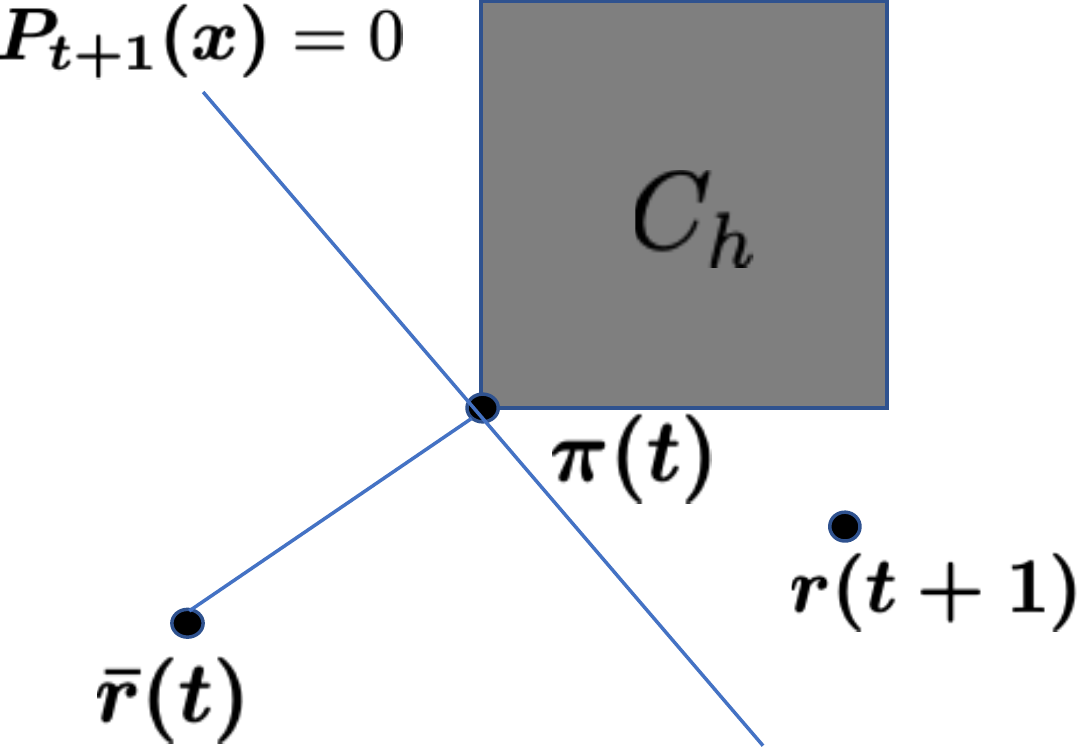}
    \caption{$\boldsymbol{\bar{r}(t)}$ and $\boldsymbol{r(t+1)}$ lie on different sides of the hyperplane $P_{t+1}$}
    \label{fig:blackwell2}
\end{figure}

{\bf Proof:}
We show that following $f_t^{\textit{F2S}}$ allocation, the average honest fraction vector $\boldsymbol{\bar{r}(t)}$ approaches the convex set $C_{\gamma}$ in $\mathbb{R}^K$ defined as follows:
\begin{align}
    C_{\gamma} &= [\gamma,1]^K \quad x \in [0,1] 
\end{align} 
so     $\boldsymbol{\bar{r}(t)} \in C_\gamma$ implies $\min_i \bar{r}_i(t) \geq \gamma$. 
Let $\boldsymbol{\pi(t)}$ denote the Euclidean projection of $\boldsymbol{\bar{r}(t)}$ to the convex set $C_{\gamma}$. Let $\boldsymbol{P_{t+1}}(\boldsymbol{x})$ denote the hyperplane perpendicular to $\boldsymbol{\pi(t)} - \boldsymbol{\bar{r}(t)}$ and containing $\boldsymbol{\pi(t)}$. Observe that $\boldsymbol{\pi(t)} - \boldsymbol{\bar{r}(t)} = \boldsymbol{u(t)}$: \begin{align}
    \boldsymbol{P_{t+1}}(\boldsymbol{x}) &: \boldsymbol{u(t)}.\boldsymbol{x} - \gamma\sum_{i=1}^K u_i(t) = 0.
\end{align}
We show using the strategy space inequality in Appendix \ref{sec:proof_ss1} (\eqref{eqn:ssp}, set $s=K$), the following:
\begin{align}
    \boldsymbol{u(t)}\cdot \boldsymbol{r(t+1)} - \gamma\sum_{i=1}^K u_i(t) &\geq 0 \; \quad \forall \boldsymbol{\beta(t+1)} \label{eq:hyperplane-sep} \\
    \boldsymbol{u(t)}\cdot\boldsymbol{\bar{r}(t)} - \gamma\sum_{i=1}^K u_i(t) &\leq 0 \nonumber
\end{align}
The above two inequalities imply that $\boldsymbol{r(t+1)}$ and $\boldsymbol{\bar{r}(t)}$ lie on different sides of $\boldsymbol{P_{t+1}}$ as depicted in figure \ref{fig:blackwell2}

Let us define $d_t$ as the euclidean distance of $\boldsymbol{\bar{r}}(t)$ from the convex set $C_{\gamma}$, i.e. $d_t = || \boldsymbol{\bar{r}}(t) - \boldsymbol{\pi(t)} ||$ , $d(\boldsymbol{a,b}) = ||\boldsymbol{a} - \boldsymbol{b}||$ for any $\boldsymbol{a},\boldsymbol{b} \in \mathbf{R}^K$. we show that 
\begin{align}
    d^{2}_{t+1} &= d^2(\mathbf{\bar{r}(t+1)},\boldsymbol{\pi(t+1)}) \leq d^2(\mathbf{\bar{r}(t+1)},\boldsymbol{\pi(t)}) \nonumber\\
    &= \left\lVert \mathbf{\bar{r}(t+1)} - \boldsymbol{\pi(t)} \right\rVert^2_2 \nonumber\\
    &= \left\lVert \frac{t}{t+1}\mathbf{\bar{r}(t)} + \frac{1}{t+1} \mathbf{r(t+1)} - \boldsymbol{\pi(t)} \right\rVert^2_2 \nonumber\\
    &= \left\lVert \frac{t}{t+1}(\mathbf{\bar{r}(t)} -\boldsymbol{\pi(t)}) + \frac{1}{t+1}( \mathbf{r(t+1)} - \boldsymbol{\pi(t)}) \right\rVert^2_2 \nonumber\\
    &= (\frac{t}{t+1})^2 \left\lVert  \mathbf{\bar{r}(t)} -\boldsymbol{\pi(t)} \right\rVert^2_2 + (\frac{1}{t+1})^2 \left\lVert  \mathbf{r(t+1)} - \boldsymbol{\pi(t)} \right\rVert^2_2 \nonumber\\ 
    &+ \frac{2t}{(t+1)^2}(\mathbf{\bar{r}(t)} -\boldsymbol{\pi(t)}).(\mathbf{r(t+1)} - \boldsymbol{\pi(t)}). \nonumber
\end{align}

\begin{align}
    (t+1)^2 d^2_{t+1} - t_2 d^2_t  &\leq \left\lVert  \mathbf{r(t+1)} - \boldsymbol{\pi(t)} \right\rVert^2_2 \nonumber \\
    &+ 2t\cdot((\boldsymbol{\pi(t)}-\mathbf{\bar{r}(t)} )\cdot( \boldsymbol{\pi(t)}-\mathbf{r(t+1)})).\nonumber
\end{align}

\begin{figure}
\centering
\begin{subfigure}{.25\textwidth}
  \centering
  \includegraphics[width=.99\linewidth]{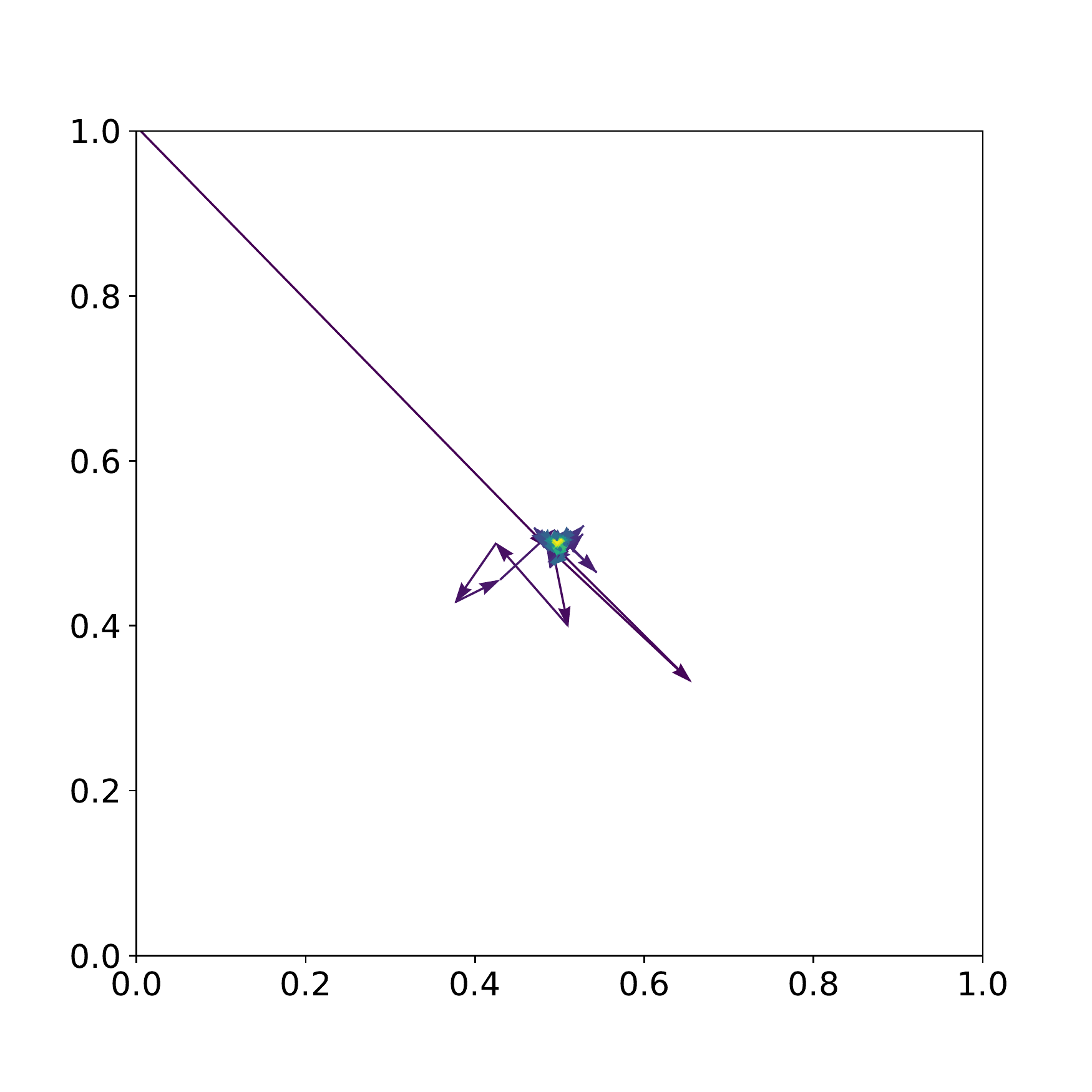}
  \vspace{-8mm}
  \caption{Convergence to $C_{0.5}$}
  %\label{fig:sub1}
\end{subfigure}%
\begin{subfigure}{.25\textwidth}
  \centering
  \includegraphics[width=.99\linewidth]{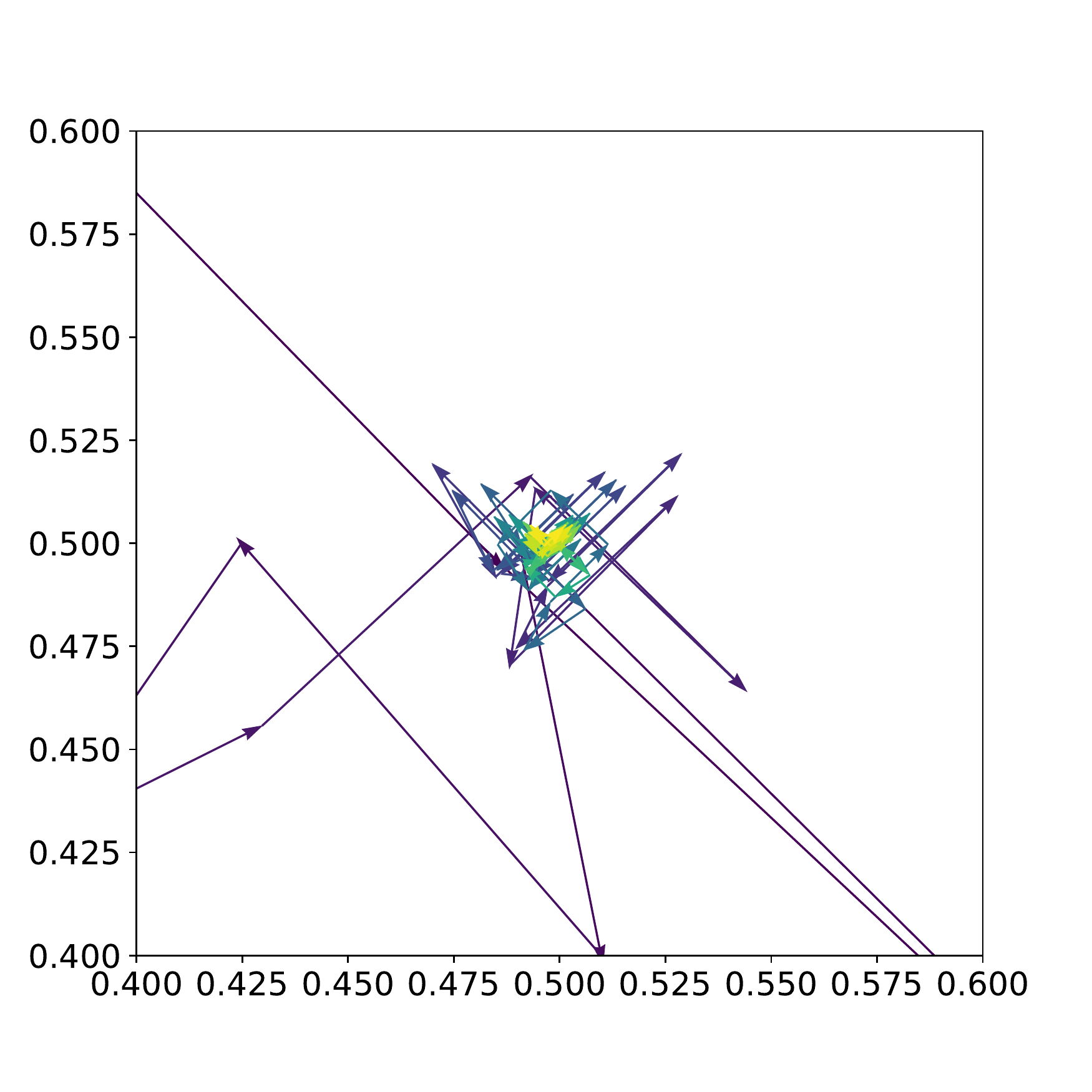}
  \vspace{-8mm}
  \caption{5x Zoom}
  %\label{fig:sub2}
\end{subfigure}
\vspace{2mm}
\caption{Evolution of ($\bar{r_1(t)}$,$\bar{r_2(t)}$), observe that the distance of ($\bar{r_1(t)}$,$\bar{r_2(t)}$) from ($0.5$,$0.5$) reduces with time (arrows signify causality)}
\label{fig:tp_evolution}
\end{figure}

$\boldsymbol{r(t+1)}$ and $\boldsymbol{\bar{r}(t)}$ lie on different sides of  $\boldsymbol{P_{t+1}}$ hence,
\begin{align}
    2t\cdot((\boldsymbol{\pi_{C_{\gamma}}(t)}-\mathbf{\bar{r}(t)} ).( \boldsymbol{\pi_{C_{\gamma}}(t)}-\mathbf{r(t+1)}))\leq 0 \nonumber
\end{align}
Moreover, 
    $\left\lVert  \mathbf{r(t+1)} - \boldsymbol{\pi_{C_{\gamma}}(t)} \right\rVert^2_2 \leq \gamma^2 K$  and combining the above two inequalities, we get:
    $(t+1)^2 d^2_{t+1} - t^2 d^2_t  \leq \gamma^2 K$. 
Summing terms over $t \in \{1,..,T-1\}$, we get
    $d_T \leq \gamma\sqrt{\frac{K}{T}}$
 and  observe that
    $d_T \geq (\gamma - \min_{i}\bar{r}_i(T))^+$. Equivalently, 
    $\min_{i}\bar{r}_i(T) \geq \gamma - d_T$ and thus 
    $\psi_{f_t^{\textit{F2S}}}(T) \leq  \gamma(1-\sqrt{\frac{K}{T}})$. 

The convergence of $\bar{r}_i(t)$ for 2 worst performing shards is illustrated in figure \ref{fig:tp_evolution}. 

\subsection{Distributed Dynamic Self Allocation}
\label{sec:ddsa}
Thus far we have studied the dynamic self allocation policy in the lens of the {\em mean} fraction of the honest nodes. What we really need is a (randomized) DSA policy that can be run by each node locally.  A natural strategy is the following: each honest node uses the \tfp DSA policy to calculate the honest node fractions for each shard and then allocates itself to one of the shards randomly, proportional to the fractions prescribed by the \tfp policy. 
While this strategy is natural and performs well in experiments (cf.\ Section~\ref{sec:simulations}) we have not been able to formally evaluate its theoretical performance. A slightly modified strategy, described below,  does enable a theoretical evaluation. 

\tfpnosp-dist DSA policy diverges from \tfp DSA policy in the following way:
The policy aims to achieve a honest node fraction in each shard of $h$, strictly smaller than the information theoretic optimal value of $\gamma$. 
With a slight abuse of notation, we define  $u_i(t-1)=(h-\bar{r}_i(t-1))+$.  The algorithm  allows honest nodes to focus on $s$ out of $K$ shards at a time; we order the quantities $u_1(t-1),\ldots, u_K(t-1)$ and define 
 $\tilde{u_i}(t-1) = u_i(t-1)$ if the index $i$ is in the highest $s$ order statistics. For other indices $i$, we set  $\tilde{u_i}(t-1) = 0$. 
 We follow the \tfp policy by substituting $\tilde{u_i}(t-1)$ in place of $u_i(t-1)$ in \eqref{eqn:main_policy}, so $\tilde{\gamma}_i = \gamma \frac{\tilde{u}_i}{\sum_{i=1}^K \tilde{u}_i}$.  Moreover, to ensure that each of the $s$ prioritized shards get some honest nodes, we lower bound their prescribed honest policy to $q/(1+2q)s$, for some constant $q$ close to 0.
We ensure this by projecting the non-zero prescribed honest power $\tilde{\gamma}_i(t)$ to the set $C_{q/s}$;  we use the notation ${\rm Proj}(\boldsymbol{\tilde{\gamma}(t)},C_{q/s})$ to denote such Euclidean projection. In summary, 
\tfpnosp-dist DSA policy is the following:
\begin{align}
    f_t^{\textit{F2S-dist}}(h,q,s): \boldsymbol{\gamma(t)} &= \frac{1}{1+q/\gamma} \quad {\rm Proj}(\boldsymbol{\tilde{\gamma}(t)},C_{q/s}).  \label{eq:trip} 
\end{align}
Notice that the honest node fraction is no longer deterministic since the honest node allocation is randomized (and follows a multinomial distribution);  each shard's marginal distribution of honest  power is  
    $\Gamma_i(t)$ is distributed as  $\frac{1}{N}{
    \rm Binomial}(n,\gamma_i(t))$ and each shard's marginal distribution of honest node fraction  
    $r_i(t)$ is distributed as  $\frac{\Gamma_i(t)}{\Gamma_i(t)+\beta_i(t)}$. 
Notice that $\psi_{f_t^{\textit{F2S-dist}}}$ is now a random variable, and we  show a concentration bound below; the proof is deferred to Appendix~\ref{app:proof_trip}. This result shows that with high probability the information theoretic upper bound of $\gamma$ can be achieved by the appropriate honest policy. 

\begin{theorem}
    For any adversarial strategy, with probability $1-\delta$, 
    \begin{align}
        \psi_{f_t^{\textit{F2S-dist}}(h,q,s)}(T) \geq \gamma\left(\frac{h}{\gamma} - \frac{1}{\gamma}\sqrt{h^2\frac{K}{T} + 4hs\sqrt{\frac{2}{T}\log \frac{2}{\delta}}}\right). 
    \end{align}
    \label{thm:tfpproj}
\end{theorem}
We note that $h=(1 - s e^{-n\frac{q}{(1+2q)s}(-c+clogc+1)})\frac{cs}{K(1-2q)}\gamma$ can be set close to $\gamma$ by choosing the variables $q,s,c$ appropriately.

\noindent {\bf Outline of Proof}. 
Following the same pattern as the proof of \tfp DSA, we show that the average honest fraction vector $\boldsymbol{\bar{r}(t)}$ approaches the convex set $C_h$.

We first prove a \textit{strategy space inequality} which states the following:
\begin{eqnarray}
 \max_{\mathbf{\gamma}} \min_{\mathbf{\beta}} \sum_{i=1}^{K} u_i(t-1) \frac{\tilde{\gamma}_i(t)}{\tilde{\gamma}_i(t)+\beta_i(t)} \geq \gamma \frac{s}{K} \sum_{i=1}^{K} u_i(t-1) \label{eqn:ssi1}
\end{eqnarray}

We then modify the target allocation policy as $\boldsymbol{\gamma}$ which ensures that $\gamma_i\geq \frac{b}{s}$ as shown in \eqref{eq:trip} where $b$ is a constant and $q = b/(1-2b)$. The new allocation policy leads to a \textit{modified strategy space inequality} given by:
\begin{align}
    \min_{\boldsymbol{\beta}} \sum_i u_i(t-1) \frac{\gamma_i(t)}{\gamma_i(t)+\beta_i(t)}  &\geq \frac{s}{K} \gamma(1-2b)   \sum_{i=1}^K u_i(t-1)
\end{align}
We now show that if every honest node chooses a shard randomly according to a choice distribution given by $\boldsymbol{\gamma}$ defined in  \eqref{eq:trip}, we get a \textit{stochastic strategy space inequality} given by:
\begin{align}
    \mathbb{E}_{\boldsymbol{\Gamma(t)}}&\left[\min_{\boldsymbol{\beta(t)}} \boldsymbol{u(t-1)}.\boldsymbol{r(t)} \right] \geq h\sum_{i=1}^K u_i,
\end{align}
%is illustrated in  Figure \ref{fig:blackwell1} where we see that the expected honest node fraction for the next time-step will lie within a space separated by the hyperplane passing through the projection on  $C_h$ and containing $C_h$. This is 
similar to \eqref{eq:hyperplane-sep} in \tfp DSA proof.
We then show that following the above honest allocation strategy time averaged honest fraction approaches the convex set $C_h$ with distance decreasing with time $T$ as:
\begin{align}
     d^2_{T} &\leq h^2\frac{K}{T} + \frac{2}{T}\sum_{t=1}^{T-1}\frac{t}{T}(Y_t) \label{eqn:blackwell_distance}
\end{align}
where  $Y_t = (\mathbb{E}_{\Gamma_i(t)}\left[\sum_i u_i(t-1) r_i(t)\right] - ((\boldsymbol{u(t-1)})\cdot\mathbf{r(t)})$ is a martingale difference sequence with respect to the  history at time $t$ and $|Y_t|\leq 2 h s$.  Using the Azuma- Hoeffding inequality, we have
\begin{align}
    \mathbb{P}\left( \frac{1}{T} \left\lVert \sum_{t=1}^{T-1} Y_t \right\rVert > \epsilon_m \right) \leq 2 e^{ -\frac{T\epsilon_m^2}{8h^2s^2}}. \label{eqn:blackwell_azuma}
\end{align}
This allows us to conclude that with probability $(1-\delta)$, the distance to convex set converges to 0 as 
$d^2_{T} \leq h^2\frac{K}{T} + 4hs\sqrt{\frac{2}{T}\log(\frac{2}{\delta})}$. We observe that
$d_T \geq (\gamma - \min_{i}\bar{r}_i(T))^+$. Equivalently, 
$\min_{i}\bar{r}_i(T) \geq \gamma - d_T$ and thus 
$\psi_{f_t^{\textit{F2S-dist}}(h,q,s)}(T) \geq \gamma\left(\frac{h}{\gamma} - \frac{1}{\gamma}\sqrt{h^2\frac{K}{T} + 4hs\sqrt{\frac{2}{T}\log \frac{2}{\delta}}}\right)$.

\subsection{Number of Shards and Nodes}
Conventional modeling (and the corresponding  sharding literature) supposes that the number of nodes $N$ is much larger than the number of shards $K$. This modeling is central to the working of  node to shard (N2S)  allocations: this way each shard has a sufficient number of honest nodes. In practice, one can imagine several shards being inactive during certain periods of time and conceivably $K > N$.
In this scenario, we can derive a tighter information theoretic  bound than the one in \eqref{eq:info_theoretic_1}  since $\psi(T) = \gamma$ implies that each shard remains active at all rounds even if there aren't sufficient honest nodes to maintain all the shards in any round; this is done next. 

\noindent {\bf Information Theoretic Limit}
The sum of honest nodes of all shards is limited by the total number of honest nodes in the system, since at every round, there will be at most $N$ shards which can be maintained by honest nodes, and the adversary can set it's policy: $B_i(t,\Gamma_i(t))=\frac{\beta}{\gamma}\Gamma_i(t)$. Thus, the honest fraction of the $N$ out of $K$ shards which are non zero is $\gamma$, yielding the following bounds:
\begin{align}
   \max_{\{f_t\}_t} \, \min_{\{\beta_i(t)\}_{i,t}} \,& \sum_i \left\{  \bar{r}_i(T) \right\} \leq \gamma N \label{eqn:finiteNSum} \\
     \max_{\{f_t\}_t} \, \min_{\{\beta_i(t)\}_{i,t}} \, \min_i \bar{r}_i(T) &\leq \frac{1}{K}\sum_i \bar{r}_i(T) \nonumber \\
    \psi(T) \leq \gamma N/K. \label{eq:info_theoretic_2}
\end{align} 
We note that the \tfpnosp-dist self allocation strategy smoothly meets this new upper bound; this is done  via the honest nodes focusing only on a subset of shards in a round to achieve $\psi_{f_t^{\textit{F2S-dist}}(h,q,s)}(T)$ which is within a $O(\frac{1}{\log N})$ multiplicative factor of the improved information theoretic upper bound: 
Theorem \ref{thm:tfpproj} states that for large enough $T$, $\psi_{f_t^{\textit{F2S-dist}}(h,q,s)}(T)\geq 0.5h$. We observe that $h \geq \frac{a}{\log N} \frac{N}{K}$ 
where $a$ depends on the choice of $c,q$ and $s = \frac{N}{4\log N}$. 
Thus,  $\psi_{f_t^{\textit{F2S-dist}}(h,q,s)}(T)\geq \frac{a}{2\log N} \frac{N}{K} \gamma$  which is within $O(\frac{1}{\log N})$ multiplicative factor of the information theoretic limit in \eqref{eq:info_theoretic_2}.

\subsection{Experiments}
\label{sec:simulations}
We empirically verify the robustness of our protocol against various randomness associated with a practical implementation such as the (theoretically accounted for) randomness in honest node allocation and the bias in estimation of adversarial and honest distributions. Our implementation consists of $K=100$ shards with varying number of nodes $N$ and heterogenous target honest node fractions. %We also propose a conjecture stating that running the DSA engine with a pure Blackwell policy ($\gamma_i = \gamma u_i/\sum_i u_i)$ will also produce the desired results. 
This experiment models  a realistic setting where the honest nodes do not know the honest fraction of all shards, have to estimate the honest node fraction in the shards by calculating the net  power (i.e., total number of nodes) and the estimated honest node allocation. 
%Our experiments show that even a biased estimator of honest node fraction given by $\frac{N\gamma_i(t)}{\mbox{Observed total shard power}}$, still yields the same performance.
We set  $\beta=0.5$ in all our experiments, the largest possible adversarial power that can be tolerated for security of the consensus engine (and the overall \tfp architecture). 

\paragraph{Adversarial strategy}.  The adversarial action is worst-case and cannot be empirically evaluated, so we identify an adversary that has the following capabilities. It can throttle the honest node fraction in any shard to be as small as $\frac{\log K}{K}$ for as long as $\tau = \frac{\log K}{\log \log K}$. We state this formally, deferring the proof to Appendix~\ref{app:latencyboundedadversary}.
\begin{proposition}
\label{prop:advlatency}
Consider an adversary that works in periods of length $\tau$. In any period at a  round $t$, the adversary uniformly allocates its power on   the $K \left (\frac{1}{\log K}\right )^t$  worst performing shards. The period ends at $t = \tau$ and the adversary restarts with $t=1$.  

Then for any honest policy, the worst case shard's honest node fraction is  less than $O(\frac{\log K}{K})$ for all rounds $t \leq  \tau$. 
\end{proposition}

\begin{figure}
\centering
\begin{subfigure}{.25\textwidth}
  \centering
  \includegraphics[width=.99\linewidth]{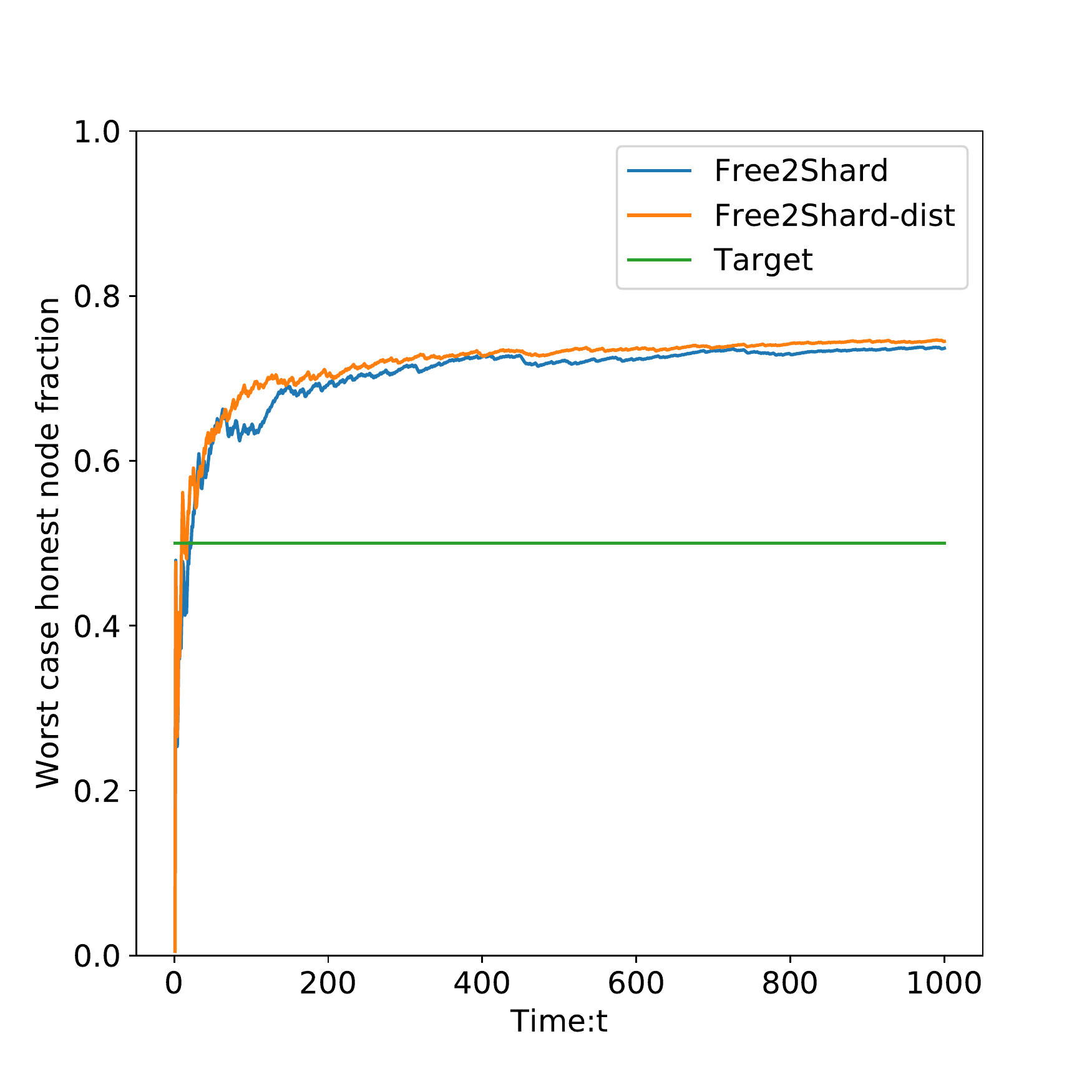}
  \vspace{-8mm}
  \caption{$N=10K$}
  \label{fig:homon>k}
\end{subfigure}%
\begin{subfigure}{.25\textwidth}
  \centering
  \includegraphics[width=.99\linewidth]{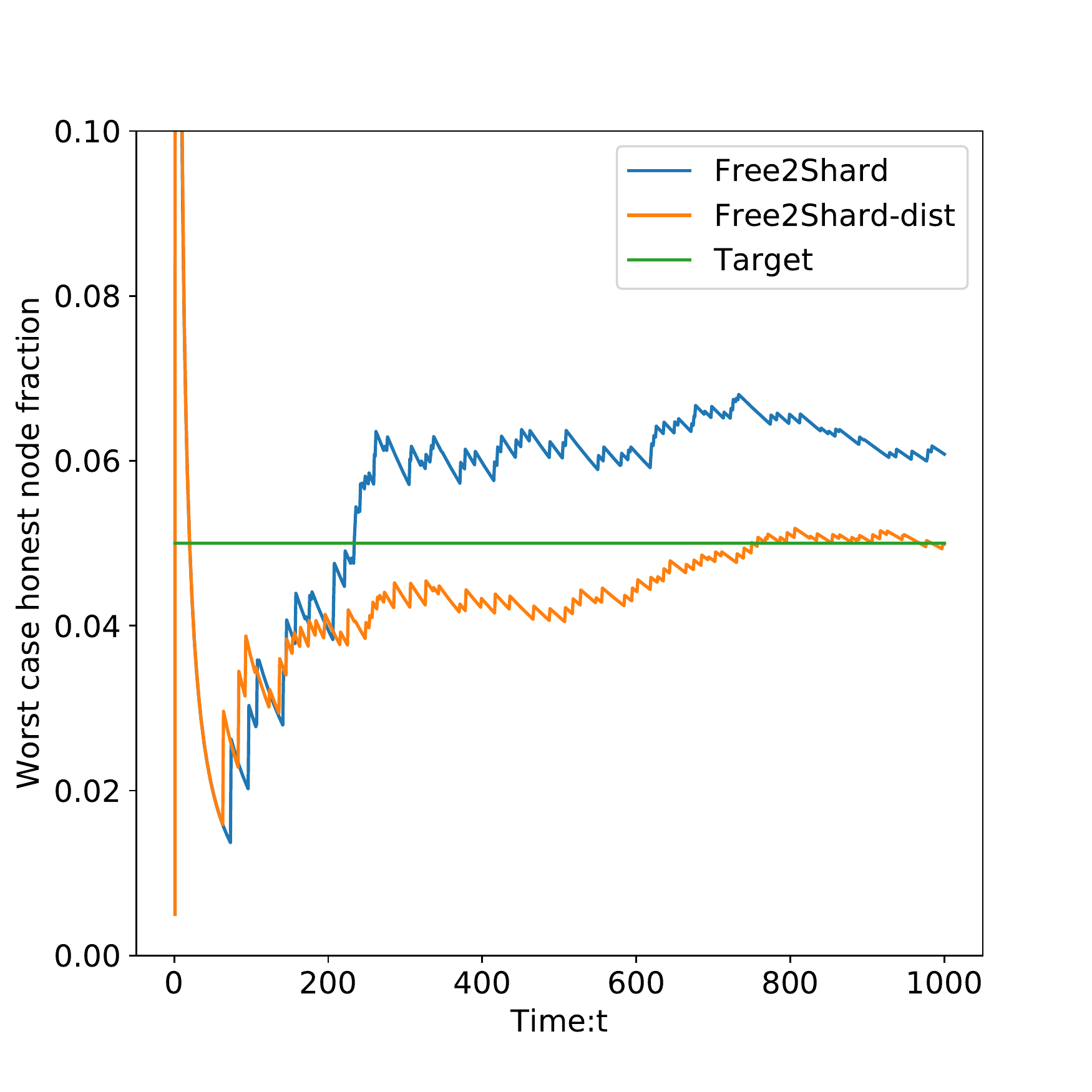}
  \vspace{-8mm}
  \caption{$N=0.1K$}
  \label{fig:homok>n}
\end{subfigure}
\vspace{2mm}
\caption{Homogeneous sharding comparison between \tfp and \tfp-dist}
\label{fig:homogeneous}
\end{figure}

\noindent {\bf Homogeneous sharding with $N>K$}. 
With $K=100, N=1000, \beta=0.5$ we simulate the adversarial strategy above and the distributed (randomized) versions of \tfp and \tfpnosp-dist as the honest policy. In this homogeneous setting, we aim to maintain the same honest node fraction in each of the shards (equal to 0.5 here). We plot the worst time average honest node fraction across the shards as a function of time in Figure~\ref{fig:homon>k} and make the following observations. (a) \tfp and \tfpnosp-dist have similar performances and favorable to the theoretical upper bound (derived for the worst-case adversary); (b) since $N$ is sufficiently larger than $K$, there are enough honest nodes to allocate to each shard and thus the honest node fraction in each shard is able to meet the target.   

%\begin{figure}
%    \centering
%    \includegraphics[width = 0.35\textwidth]{images/latency_bound_adv/compare_worst_shard_tp12.pdf}
%    \put(-180,45){\rotatebox{90}{Worst case honest fraction }}
%	\put(-100,5){Time: $t$}
%    \caption{Homogeneous sharding: $N=10K$ }
%    \label{fig:homon>k}
%\end{figure}

\noindent {\bf Homogeneous sharding with $N<K$}. 
With  $N=10, K=100, \beta = 0.5$,  we simulate the adversarial strategy above and the distributed (randomized) versions of \tfp and \tfpnosp-proj as the honest policy. In this homogeneous setting, we aim to maintain the same honest node fraction in each of the shards -- since there are more shards than nodes, the largest possible target is  $\frac{N(1-\beta)}{K} = 0.05$.  We plot the worst time average honest node fraction across the shards as a function of time in Figure~\ref{fig:homok>n} and make the following observations. (a) \tfp and \tfpnosp-proj have similar performances and favorable to the theoretical upper bound (derived for the worst-case adversary); (b) even though $N$ is sufficiently smaller than $K$, so each node's randomized  decision cannot be expected to average the honest node fraction in each shard to the deterministic values proposed by \tfpnosp, we see that the performance in meeting the targets is surprisingly strong. A theoretical justification for this phenomenon is of great interest and is deferred to future work.

%\begin{figure}
%    \centering
%    \includegraphics[width = 0.35\textwidth]{images/latency_bound_adv/compare_worst_shard_tp35.pdf}
%    \put(-180,40){\rotatebox{90}{Worst case honest fraction }}
%	\put(-100,5){Time: $t$}
%    \caption{Homogeneous sharding: $N=0.1K$ }
%    \label{fig:homok>n}
%\end{figure}

\begin{figure}
    \centering
    \includegraphics[width = 0.3\textwidth]{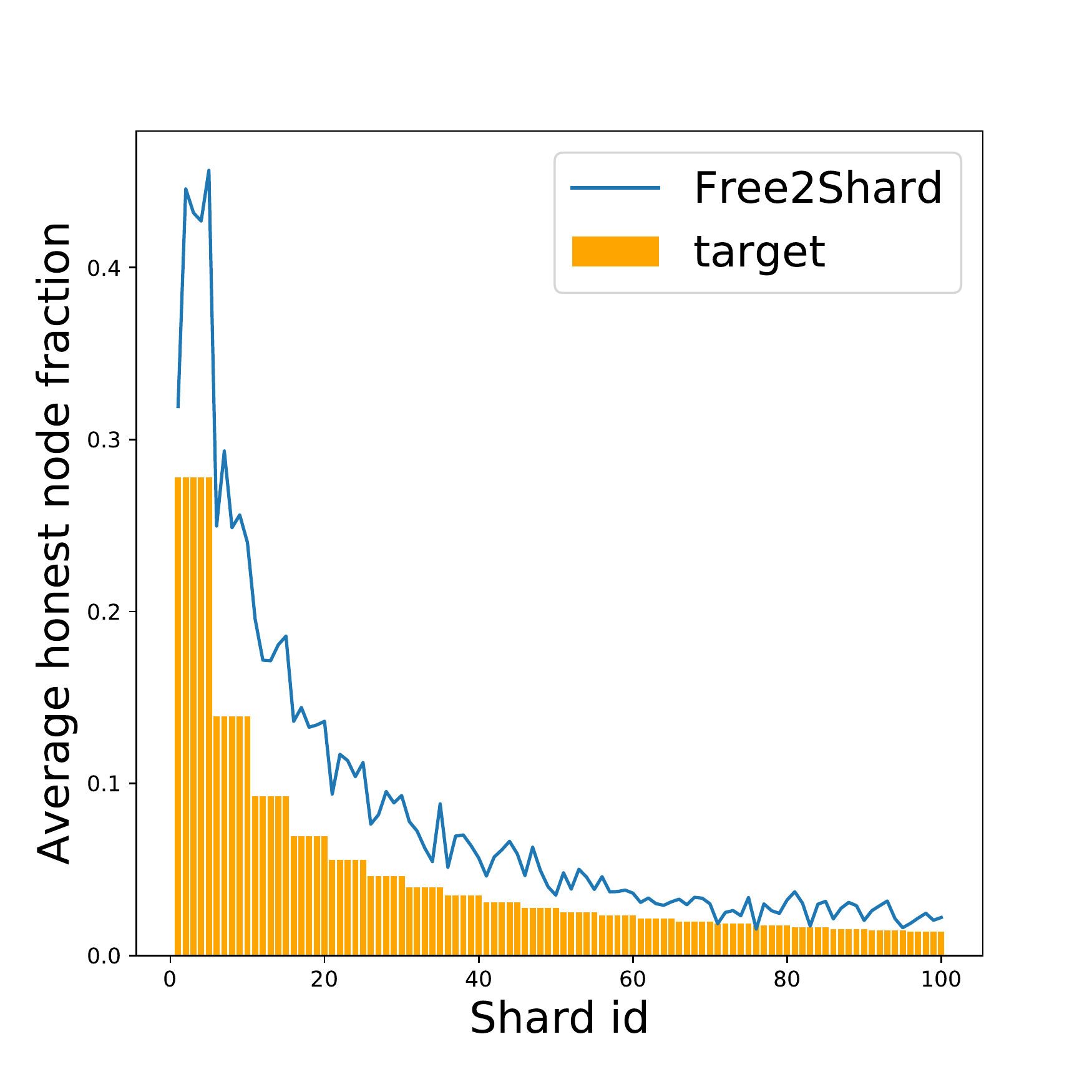}
    %\put(-180,45){\rotatebox{90}{Worst case honest fraction }}
	%\put(-100,5){Shards}
    \caption{Comparison between target honest fraction set for heterogeneous sharding and average honest fraction achieved at the end of the experiment  }
    \label{fig:hetero}
\end{figure}

\paragraph{Heterogeneous sharding with $N<K$}. 
In practice, different shards have different activity levels, demanding different target levels of participation from the nodes. We propose the target honest fraction of each shard to decrease as $1/(\left \lceil{i/5} \right \rceil+1)$, 
%(see Figure \ref{fig:hetro_target1}), 
with $N=10, K=100, \beta=0.5$; where $i$ is the index of the shard. We plot the time average honest node fraction across the shards at the end of the simulation in Figure~\ref{fig:hetero} and make the following observations. (a) \tfp supports heterogeneous target honest fraction vector allocation across shards;  (b) even though $N$ is sufficiently smaller than $K$, we can set some shards to achieve a target honest factor of $1-\beta$ which is the best we can achieve even with $N>K$. We see that the performance in meeting the targets is  strong; a theoretical justification for this strong performance for heterogeneous sharding is of great interest and is deferred to future work.

\section{\tfp Architecture Design}
\label{sec:tfp}

In this section  we discuss in detail the \tfp sharding architecture, an overview of which was provided in Section~\ref{sec:overview}.

Each node $i$ in the network has an identity which is comprised of a pair $(nsk_i,npk_i)$ the node secret key (known only to the node) and node public key (known to everyone). 
We will assume that the overall blockchain can be split into $K$ sub-ledgers (called shards) each comprising of a entirely {\em independent} set of transactions (submitted by shard clients). We can use methods developed for inter-sharding in other sharding algorithms (for example, \cite{kokoris2018omniledger}), and will briefly allude to this in Section~\ref{sec:systemView}. Each shard will have its own peer-to-peer (p2p) network in which shard transactions are broadcasted. Any node desirous of reading the activities of the shard can join the p2p network.

\tfp sharding architecture relies on a State Machine replication engine (SMR) maintained by all the peers in the network to create a total ordered log. Shard blocks are ordered using the hash pointers on the ordered log. Since the ordered log contains only hash of each shard block, data availability is ensured by invoking a separate voting mechanism based on a recent primitive, Coded Merkle Tree \cite{CMT}.   \tfpnosp-dist DSA policy ensures liveness and high throughput against adaptive adversary. Finally, the rotation of nodes to a new shard is facilitated by periodic state commitments on the ordered log. We explain each of these functionalities below and show that they have $o(1)$ overhead as the shard block size $B$ becomes large.

The adaptive-adversary resistant SMR has a mechansim for growing the ordered log based on transactions sent to it, we further assume that it has an upper-bound on the latency of including honest transactions into the ordered log (which is independent of $N$). This property holds for both the Algorand and Ouroboros Praos protocols. We will assume that the ordered log is comprised of a sequence of transactions organized into blocks (we will refer to the SMR-block-number as the latest block number in the SMR). We do not require fresh randomness from the ledger, but in practice, this can be useful and in which case instead of the SMR-block-number, we will use the randomness associated with SMR-block-number. 
We are now ready to discuss the \tfp architecture and its components in detail. 

\subsection{\tfp~components and design}
\noindent{\bf Shard block mining:} We create a ``mining'' mechanism (a permissioning mechanism) for allowing the creation of shard blocks. A node $i$ can propose a shard block for inclusion in the ordered log if $H(npk_i,SMR_{no})$ is small, where $H$ is a cryptographic hash function. A node will have to specify one of the shard-id on which wishes to mine a block, as well as the hash of the shard block.  The SMR engine collects all the hash outputs from the nodes, and selects the smallest $\kappa$(a constant) output values per shard to propose blocks. The shard-hashes enter the ordered log. Note that each peer can only generate at most one shard-block per SMR-block. This ensures that even adversarial nodes have to make a choice on which shard it is mining.

The above mining protocol ensures if $\Gamma_i\cdot N$ and $\beta_i \cdot N$ nodes are interested in mining in shard $i$, then, owing to the uniform nature of the hash output string, the average fraction of honest shard blocks mined is $ \Gamma_i/(\Gamma_i+\beta_i)$, and consistent with the quantity studied in Section \ref{sec:dsa}. The mining mechanism thus ensures that if a shard has a capacity to process $R$ transactions per second it will process $ R\Gamma_i/(\Gamma_i+\beta_i)$ honest transactions per second.

While we  describe our protocol using hash functions for simplicity, in order to be resilient to an adaptive adversary, we will use verifiable random functions (VRF) \cite{vrf,vrf2}. VRF  ensures that the output of the function is not predictable by any other node {\em a priori}. We refer the reader to Section~\ref{app:vrf} for a brief description of VRF.

\noindent {\bf Shard ledger:} The ordered log of shard block hash pointers corresponding to a given shard-id induces a shard ledger comprising of the corresponding shard blocks. We note that the architecture does not guarantee validity of every transaction in the shard ledger, since SMR nodes do not checked for validity. However, this is not an issue as long as there is a consistent execution engine that will interpret the shard transactions. This is a classic idea in distributed systems of decoupling execution from ordering \cite{yin2003separating}, and has found applications in blockchain scaling architectures as well \cite{balakrishnan2013tango}.

\noindent {\bf Data Availability}:
Even though the shard blocks need not be valid, they still need to be available for retrieval. We accomplish this availability check using a mechanism by which all nodes vote through the SMR on whether a block is available. The key idea is the following: the node whose shard block hash got included in the ordered log, now sends forth small chunks of the block encoded appropriately, one for each node. The original block is divided into $K$ chunks and is coded to form $N$ chunks. Which chunk is to be sent to which node is determined by a $H(npk_i,SMR_{no})$ (a random mapping of the public key of the recipient $npk_i$ along with the SMR block number). 

Now in the SMR, each node casts its vote on whether a previous shard block pointer is valid or not, depending on whether it received its chunk. We need to establish three properties: (1) enough number of nodes received the chunk, (2) the chunk satisfied the hash of the coded block, and (3) the block was correctly coded (as an adversary may do otherwise). The first constraint can be satisfied as long as there is a majority of votes cast for validity. The $\beta<0.5$ fraction of nodes are adversarial and may lie, however, the remaining $a = 0.5-\beta$ honest nodes have indeed received the chunk. Thus as long as $aN$ chunks are sufficient to reconstruct the block using the code, property (1) is satisfied. Property (2) can be satisfied by requiring that the posted hash pointer corresponds to the coded block rather than the uncoded block. Finally, Property (3) is the most subtle property to satisfy. We  note that recent work on Coded-Merkle-Tree has solved this problem by enabling short fraud proofs (short statements of length $\Theta(\log B)$) that can prove that the block was incorrectly coded - and such proofs are readily forthcoming as long as there is enough chunks to decode a correctly coded block. A shard block, even when voted for by a majority, will be considered invalid, if some node posts a fraud proof.

\noindent {\bf State commitment:} When nodes rotate between shards, they will need to fully download the new shard ledger as well as execute it from the beginning of history in order to synchronize to that shard. This can be extremely resource intensive and can completely drown out any gain due to sharding. We point out the multi-consensus protocols such as \cite{luu2016secure,kokoris2018omniledger,zamani2018rapidchain} which assume majority-honest in each shard do not have this problem, as they can immediately assume that any previous transaction that has been signed by a majority is valid. Instead, if the node could acquire a trustworthy state after executing till the previous block, it can easily rotate into the shard. We note that one possibility is to use verifiable computing primitives to assert that the state of the shard ledger as executed by another node is correct \cite{parno2013pinocchio,ben2014succinct,ben2018scalable}, however, these results are still not yet practically applicable to general program execution, so we resort to a different mechansim for state commitments. 

State commitments consists of the merkle root of a Merkle Patricia Trie \cite{luu2017peacerelay} of a shard's execution state, which is posted at regular intervals of $E$ SMR-blocks called epochs. They serve as checkpoints for fast bootstrap and inter-shard transactions. The state commitment is posted to the SMR engine by a epoch leader who is elected every epoch. Each node locally computes {\em H($nsk_i$,$SMR_{no}$, shard-id)} for all $K$ shards. If for shard $i$, the output of the hash is less than a {\em threshold}, the node assigns itself as a epoch leader for the next epoch in that shard. The {\em threshold} is set globally to ensure that roughly a constant number of nodes are always elected as epoch leaders per shard. 

\begin{figure}
    \centering
    \includegraphics[width = .45\textwidth]{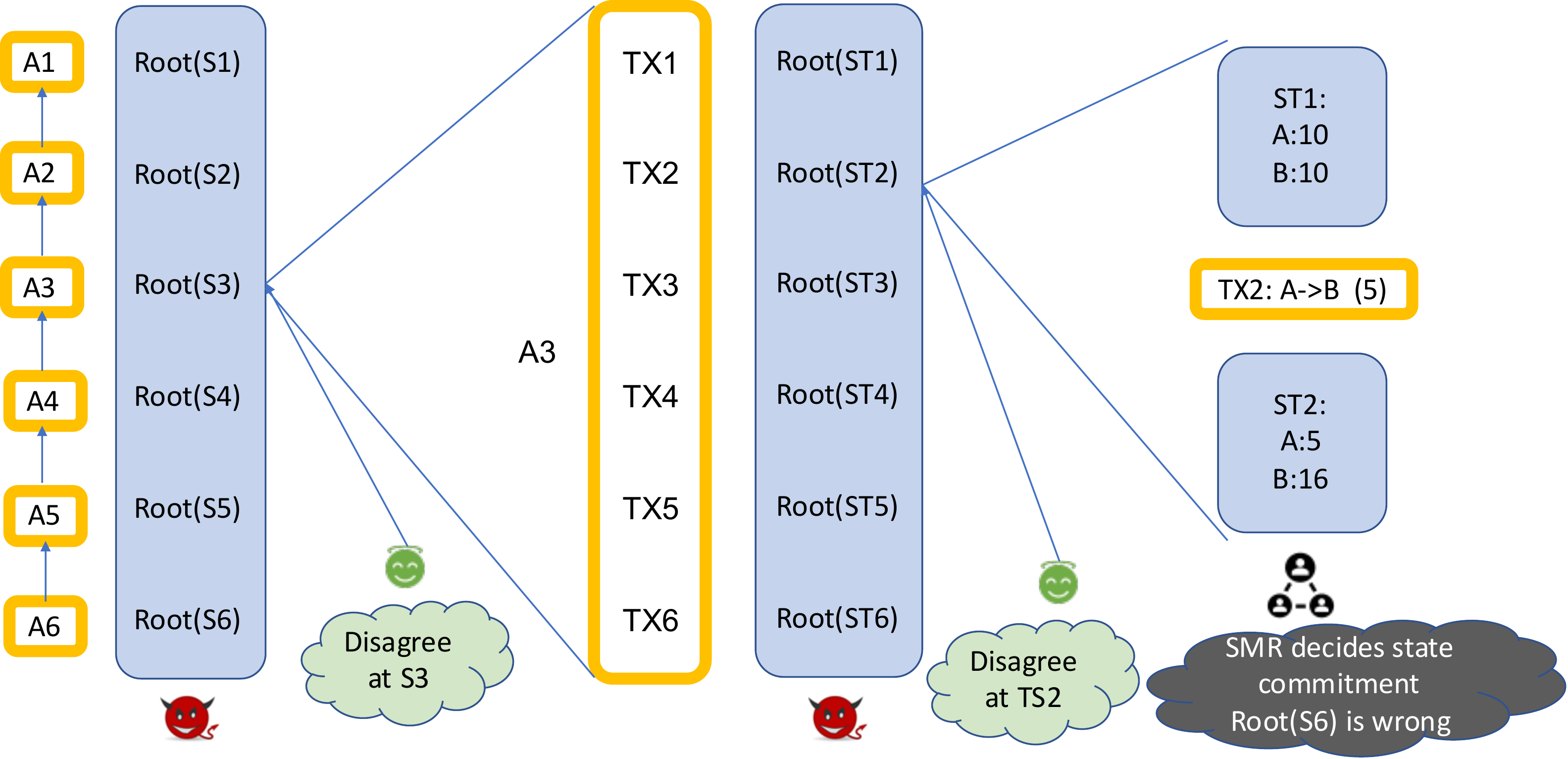}
    \caption{State Commitment challenge}
    \label{fig:challenge}
\end{figure}

The state commitment is termed invalid if it corresponds to an incorrect execution state. We can assume that epoch leaders can be byzantine and post invalid state commitments. We use a round based interactive fraud proof protocol inspired from Truebit \cite{truebit} and Arbitrum \cite{kalodner2018arbitrum} to detect such invalid commitments (we refer the reader to \cite{truebit} for a detailed explanation of the protocol). An epoch is split into $R$ rounds with each round consisting of $G$ SMR blocks. 
The interactions between the epoch leader and a challenger is illustrated in Figure \ref{fig:challenge} can be briefly described as follows:
\begin{enumerate}
    \item The epoch leader posts $S_s$ intermediate states for the challenged state.
    \item A challenger responds with a number indicating the first intermediate state when the challengers view differs from the leader
    \item The game continues to the next round with the leader posting intermediate states for the smaller challenged state and a new challenger responding according to 2.
\end{enumerate}

The game ends when the conflict is resolved down to one transaction and that transaction is posted on the SMR. We note that the number of rounds is $\Theta(\log B)$.

Note that any node can be a challenger to exactly one shard of its choice in a round, by doing so, it elects itself to the challenge committee of that shard, however it is free to join any other shard's challenge committee in the next round. The dynamic challenge committee makes the challenge protocol robust to an adaptive adversary and we can claim 
that the state commitment protocol satisfies validity and liveness:
\begin{itemize}
    \item \textit{Validity:} A state commitment corresponds to a valid state if every challenge committee has at least one honest node, which will not be violated w.h.p. since $\gamma_i\geq q/(1+2q)K$ (the expected number of honest nodes is $Nq/(1+2q)K$ which is $\Omega(\log K)$).
    \item \textit{Liveness:} A valid state commitment will be posted in expected constant number of epochs.
\end{itemize}

\subsection{ Data Structures and Resource Consumption}
\label{sec:resources}

Let us define the {\em overhead ratio} as the ratio of (resource used to maintain the SMR) to (resource used to maintain any one shard). We define resource usage as a  3-dimensional vector: (1) Computation (2) Communication and (3) Storage. We will compute the overhead as the maximum value of the $3$-dimensions. Here, we account for resources consumed by all data structures involved in maintaining the total ordered log; this is discussed in detail next.  % and show first that {\em overhead ratio} is $(o(1),o(1),o(1))$ in all three dimensions as the block size $B$ becomes large. %Resource used throughout this section corresponds to the average resource used per shard block in the node's shard.

\noindent {\bf Total ordered log}: The ordered log is maintained by the SMR and consists of the following entries:
\begin{itemize}
    \item \textbf{Shard block pointers:} Consists of the tuple (hash of the shard block, shard-id, mining proof). $K$ such shard block pointers are appended to the log for every shard block and each pointer is of size $\log K$. Resource usage = $O(K\log K)$. 
    \item \textbf{ Shard State commitments:} There are $O(K)$ state commitments per epoch. We set the epoch duration $E$ such that in expectation, there is one shard block per epoch. Resource usage = $O(K)$. 
    \item \textbf{ State commitment challenge interactions:} Consists of challenges and replies from epoch leaders. There are $O(N)$ challenges in total which accounts for $R=O(\log B)$ interactions. Resource usage = $O(N\log B)$.
    \item \textbf{ Data availability votes:} {\em Signed} (Availability, shard block hash). Constant size posted for $K$ shard blocks and needs $O(N)$ votes per shard block.
    Resource usage =  $O(NK)$.
    \item \textbf{ Data availability fraud proofs:} Incorrect coding proofs of size $O(\log B)$ for at most $K$ shards.
    Resource usage = $O(K\log B)$.
\end{itemize}

%\noindent{\bf Shard ledger}: A node downloads all shard blocks of its shard ($i$) and orders them according to the order log. The peer goes through the order log and filters shard pointers with shard id $i$. The shard blocks are ordered according to the filtered pointers and these ordered shard blocks create a shard ledger. The shard ledger is sanitized to remove invalid transactions. The resource used shard ledger maintenance is $O(B)$ with $0$ overhead since maintaining a shard ledger is the primary aim of a node.

{\bf Shard rotation:} A node rotating to a new shard needs to synchronize to the state of the new shard. The synchronization involves downloading the state corresponding to the latest state commitment and processing shard blocks proposed after the latest state commitment. The resource cost is $O(1 + B)$ per new allocation. We set the rotation interval every $T_r$ blocks for the \tfp DSA policy. A node also needs to synchronize to another shard if it is chosen as its leader. The epoch leader election will elect a node with probability $O(K/N)$ per epoch. We set $T_r=K$ to get the resource usage as $O(B/K + BK/N)$.

{\bf Data availability requests:} A newly mined block receives requests for $O(N)$ random chunks for the base CMT symbol of size $O(B/N)$, resource usage = $O(B \log N)$ per shard. Since each shard has $O(N/K)$ nodes, the load can be distributed equally amongst all nodes, resource usage per node = $O(\frac{BK}{N}\log N)$. Each node receives $K$ chunks of size $O(\frac{B}{N} \log N )$. Resource usage = $O(\frac{KB}{N} \log N)$.

\noindent{\bf Total resource usage}

The total resource usage per shard block is given by $O(K\log K + N\log B + NK + B/K + BK/N )$ in all 3 resource dimensions, Thus {\em Overhead-ratio} = $O \left(\frac{K\log K}{B}+\frac{N\log B}{B} + \frac{NK}{B} + \frac{K\log N}{N} \right)$ $=o(1)$, as $B$ becomes large and $N > \Omega(K \log K)$. Note that we require $N > \Omega(K \log K)$ to ensure that our DSA algorithm throughput approaches $\gamma$; this requirement is also essential to  guarantee the safety of state commitment.

\subsection{Summary}

The building blocks and their interactions and the resources consumed therein allow us to  infer the following properties of \tfpnosp. 

\begin{enumerate} 
\item  {\bf Safety} of the shard ledger directly follows from the safety of the total ordered log, maintained by the SMR engine. In this manner, the core safety property (under adaptive adversaries, cf.\ Section~\ref{sec:keyprimitives}) of the SMR engine is extended to \tfpnosp. 
\item {\bf Liveness}.  The DSA engine, along with the shard block mining mechanism, guarantees that at least $\gamma$ fraction of blocks are honest in each shard over a  long enough timescale (Theorem~\ref{thm:tfpproj}). The random dynamic rotation ensures new honest blocks are produced in each shard to maintain a consistent $\gamma$ fraction of honest blocks; this ensures liveness under adaptive adversaries. %We note that we choose the shard rotation period such that each node proposes only $1$ block in expectation, so that after the identity is revealed, corruption is not relevant.
\item {\bf Scaling}. As discussed above, the  overhead ratio %is $O(\frac{N \log B}{B}+\frac{K \log K}{B}+\frac{NK}{B})$, which 
can be made arbitrarily small by making $B$ large. The average time interval between consecutive shard blocks in a shard can increased to ensure that the shard processes $O(R)$ honest transactions per second.
%We can ensure that the throughput of each shard is constant and within its resource capacity by increasing inter-block time for shard blocks and making $B$ large.
\end{enumerate}

\section{\tfp~system view} \label{sec:systemView}

We conclude the paper by discussing a variety of practical and real-world system issues in implementing \tfp (and  sharding protocols in general) in distributed permissionless blockchains. 

%\tfp provides a modular sharding architecture that is centered around the the \tfp-dist Self allocation policy. We show that the distributed \tfp-dist policy is more secure the a centralized Node to Shard allocation algorithm as our sharding architecture is resistant to fully adaptive adversaries. Moreover we also get efficiency gains compared to the N2S allocation based protocols that require a large shard size (>100)\cite{zilliqa}. A shard size of 100 implies that each computation a shard is repeated across 100 nodes, a redundancy that may be uneconomical for several real world applications. On the other hand, \tfp is secure even if the number of nodes per shard is small, we show that it is secure even in $N<K$, i.e. number of nodes per shard is less than 1. With $N<K$, we get within a $1/\log N$ factor of the the resource constrained information theoretic limits \eqref{eq:info_theoretic_2}, i.e. we can get a security guarantees of $N$ nodes by only replicating resources $O(\log N)$ times.

{\bf Heterogeneous shard throughputs}. While  most sharding protocols focus on allocating equal resources to all shards, in practice different shards will have different throughput requirements. The \tfp architecture is unique in being able to handle arbitrary throughput requirements for different shards. In particular, consider the extreme example where  the number of shards is much larger than the number of active nodes ($N \ll K$). Existing protocols cannot operate in this regime, even though most of the shards are at low levels of activity and throughput. As we showed in Sec~\ref{sec:simulations}, the DSA algorithm can achieve optimal performance even in this regime. 

\noindent {\bf Heterogeneous resources}.  In practice, different nodes will have different amount of computation resources. Particularly in permissionless deployments, for example, in proof-of-stake, the amount of stake held by a node need not match the amount of computation power held by the node. In typical sharding protocols with a N2S allocation, the nodes are allocated proportional to the stake and not to their computation power. However in \tfp if a node has high computation power, it can participate in multiple shards and contribute state commitments as well as challenges in different shards.

%

%Moreover, it can be shown theoretically that \tfp-dist can achieve heterogeneous honest fractions by setting taking a convex combination of $\boldsymbol{\tilde{u}_i(t)}$ vector in \eqref{eqn:ssi1}.

\noindent{\bf Asynchronous rotation}.  We have assumed that the honest nodes rotate at every round $t$;  however, this may not be feasible in a realistic setting. We now argue that our results hold if honest nodes rotate only once every $\Delta$ rounds. Let $\gamma_i(t)$ = $\gamma_i[n] \; \forall t \in \{(n-1)\Delta+1, ..,n\Delta\}$ and $\beta_i[n]=\sum_{t=(n-1)\Delta+1}^{n\Delta} \frac{1}{\Delta}\beta_i(t)$ 
We observe that $\frac{\gamma_i(t)}{\gamma_i(t)+\beta_i(t)}$ is convex in $\beta_i(t)$, thus, the time-average of a given round is lower bounded as follows:
\begin{align}
    \sum_{t=(n-1)\Delta+1}^{n\Delta} \frac{\gamma_i[n]}{\gamma_i[n]+\beta_i(t)} &\geq \sum_{t=(n-1)\Delta+1}^{n\Delta} \frac{\gamma_i[n]}{\gamma_i[n]+\beta_i[n]}.
\end{align}
Thus, the optimal adversarial strategy is to modify $\boldsymbol{\beta}$ at the same rate as $\boldsymbol{\gamma}$; this shows that the previous results transalte to slower rotation dynamics as well. Moreover, the results hold when nodes randomize their rotation choice (choosing to rotate with probability $\frac{1}{\Delta}$). This makes for a distributed implementation and has the added benefit that   asynchronous rotation alleviates network load by preventing  focused communication load at the same time.

\noindent{\bf A small minority can stabilize the entire protocol}.  Not all honest nodes need to follow \tfpnosp-dist policy:  Theorem \ref{thm:tfpproj} remains true even if a small (but constant) fraction of the honest nodes follow the policy. For example, the \tfpnosp-dist policy allows us to set $\gamma=0.1$. We note that this is a very valuable practical setting, since many blockchains have their own foundations that own sub-majority stake. In contrast to  existing protocols, even a small minority following the rotation policy can stabilize the throughput of all shards in \tfpnosp.

%For \tfpnosp-dist policy calculations, we can assume that the small fraction of honest nodes following the protocol is $\gamma$ and the rest all honest nodes contribute to $\beta$. 

\noindent{\bf Inter-shard transactions:} While we previously only discussed intra-shard transactions, a large class of applications require inter-shard support. State commitments can be used to facilitate inter-shard transactions. Since a state commitment satisfies validity and liveness, we can directly use existing inter-shard transaction protocols like Atomix \cite{kokoris2018omniledger}. A recent work \cite{sonnino2019replay} reported that these existing protocols are susceptible to fatal synchronization attacks and message replay attacks, we note that \tfp is not susceptible to these attacks due to global ordering. Inter-shard transactions using this method require a latency larger than shard-commitment duration. However,  since there is a global ordering of transactions, computationally powerful nodes can maintain a view of multiple shard chains and certify an inter-shard transaction for a fast inter-shard swap (earning some rewards on the way).

\noindent{\bf Incentives}. Rational nodes are incentivized to follow the DSA policy since the transaction fees on shards with lower than desired throughput will be higher due to a high supply-demand gap. Moreover, state commitment and data availability fraud proofs can require deposits that can be snatched under clearly provable malicious behavior. Thus nodes will be incentivized further to follow the protocol. 

\noindent{\bf Permissionless setting}. 
An obvious extension of \tfp is to permissionless proof-of-stake (PoS) systems. Here all that really needs to be done is to replace the underlying SMR with a proof-of-stake system like Algorand \cite{gilad2017algorand} rather than a permissioned system. Finally, we consider the adaptation of \tfp to proof-of-work (PoW). A natural approach is  to elect a committee of PoW mining participants from within the part of the blockchain that has reached consensus and this committee serves as the participants of the permissioned \tfp architecture. One way to achieve this is via  the hybrid consensus approach \cite{hybrid}: the committee consists of a set of successive miners that are deep enough in the longest chain. However, this approach is not safe against an adaptive adversary - who can clearly bribe the set of chosen miners. A problem with simply applying \tfp to this setting is the issue that miners may arbitrarily decide to attack a given shard. We can solve this problem here by using a hash-sortition between the SMR engine and the shard block mining process - this idea is adapted from \cite{prismtheory} and was originally used in \cite{backbone,fruitchains}. The key properties of \tfp all translate to this setting as well. 

% Points already in \tfp advantage:
% 1) No need for node identity management or node-to-shard.
% 2) Each group can be very small (no necessity for majority honest ).
% 3) Can handle N strictly smaller than K
% 4) Can handle heterogeneous shard throughputs.

% Further things to talk about:
% 0) Can be integrated into PoW. This is \tfp-longest-chain. Borrow material 
% from \tfp.tex here. 
% 1) No focussed loading -> (Periodic rotation) -> No need to have synchronous rotation
% Can adaptively decide which challenges to participate in.
% 2) A small minority can stabilize the entire protocol. Only this minority needs to follow rotation. Only this minority needs to participate in challenge protocol. 

% 3) Heterogeneous situation: Scenario where the amount of stake / mining power controlled by a node is very different from the computation power contributed by the node. 
% -> Foundation node. small amount of stake but has large compute power.
% 4) Inter-sharding: Direct application of state commits.
% -> give generous to Atomix in Omniledger.
%  - Fast inter-sharding: If you know the state of the other protocol, can under-write transactions for fast swap.
%  - Global order => If you have computation power, which inter-shard transactions are valid => You can certify inter-shard transactions easily
% 5) Incentives: 1) The transaction fee in a shard will raise if under-provisioned. 2) We can also charge fee for writing shard data into the SMR 3) We can have state commitments and data availability claims have deposit.

%\input{implementation}
%\input{conclusion}
%\input{proofs}

\bibliographystyle{ACM-Reference-Format}
\bibliography{bib}

% % --- Appendix ---%
\appendix
%\input{general}
%\section{Appendix}
%\label{sec:set-diff-dodis}
\section{Proofs}

\label{sec:proof}

\subsection{Proof of Proposition~\ref{thm:dsa1}}
\label{sec:proof_dsaa}
\begin{proof}
    \textbf{Lower bound:} \\
    Let $a_t := \beta_1(t)$, $b_t = \frac{\gamma}{2\beta} a_{t} + \frac{\gamma}{2K}$, $l = \frac{\gamma}{2K}$, $u=\frac{\gamma}{2} + \frac{\gamma}{2K}, \omega = \frac{2\beta}{\gamma}$. We have $\frac{\gamma}{2K} \leq b_t \leq \frac{\gamma}{2} + \frac{\gamma}{2K}$.
    \begin{align}
        \psi(K) &=  \min_{\{\frac{\gamma}{2K} \leq b_t \leq \frac{\gamma}{2} + \frac{\gamma}{2K} \}}    \quad  \left\{ \frac{1}{T} \sum_{t=1}^T \frac{  b_{t-1}}{   b_{t-1}+(\frac{2\beta}{\gamma}b_t-\frac{\beta}{K})} \right\} \nonumber \\
        &\geq  \min_{\{\frac{\gamma}{2K} \leq b_t \leq \frac{\gamma}{2} + \frac{\gamma}{2K} \}}    \quad  \left\{ \frac{1}{T} \sum_{t=1}^T \frac{b_{t-1}}{  b_{t-1}+ \frac{2\beta}{\gamma} b_t} \right\} =: \phi(K) \\
        \phi(K) &= \min_{\{l  \leq b_t \leq u \}}    \quad  \left\{ \frac{1}{T} \sum_{t=1}^T \frac{b_{t-1}}{  b_{t-1}+ \omega b_t} \right\}    \label{eqn:optimum0} 
    \end{align}
    Now, we call a time-instant $t$ $g$-good if the throughput at that time is greater than $g$. We will show that the adversary does not have the ability to have a consecutive run of $g$-bad instances for more than $\tau$-periods. 
    Suppose an instant $t$ is not $g$-good. 
    Then 
    \begin{eqnarray}
    \frac{b_{t-1}}{  b_{t-1}+ \omega b_t} \leq g \nonumber 
    \end{eqnarray}
     which implies ${b_t} \geq b_{t-1} r$ where   $r = \frac{1-g}{\omega g}$. 
    Note $g \leq \frac{1}{1+\omega}$ implies $r \geq 1$. Suppose there are $\tau$ consecutive $g$-good instances at time $t$.
    Then ${b_t} \geq b_{t-\tau} r^{\tau}$. Given $b_{t} \leq u$ and $b_{t-{\tau}} \geq l$, we have $u \geq l r^{\tau}$. This implies 
 $tau \leq \log_r \left( 
    \frac{u}{l} \right)$. 
    Thus there is a $g$-good instant every $\tau+1$ instances. This implies that the throughput is at least
    $\frac{g}{\tau+1}$.
    \begin{eqnarray} \phi(K) \geq  \frac{g \log r}{\log \left( \frac{ur}{l} \right)} \geq \frac{g \log \left( \frac{1-g}{\omega g} \right)}{\log \left(r  \frac{\frac{\gamma}{2}+\frac{\gamma}{2K}}{\frac{\gamma}{2K}} \right)}. \nonumber 
    \end{eqnarray}
    Consider a worst case scenario where $\beta=0.5$, $\gamma=0.5$,
    for all $g \leq \frac{1}{1+\omega}= 1/3$,  choosing $g=1/4$, and  $1/2+1/2K \leq 1$, we get: 
    \begin{equation}
    \phi(K) \geq \frac{0.14}{\log_2(3K)}. 
    \end{equation}
    
    \textbf{Upper Bound:}
    We demonstrate an adversarial strategy which holds to the following claim: $\psi(K) \leq O (\frac{\log \log K}{\log K})$.
    Compare the denominators of $\psi(K)$ and $\phi(K)$: 

    \begin{align}
    { \frac{\gamma}{6\beta} a_{t-1} + \frac{\gamma}{6K} +  \frac{\gamma a_t}{3\beta} + \frac{\gamma}{3K}} &\leq {\{ \frac{\gamma}{2\beta} a_{t-1} + \frac{\gamma}{2K} \}+a_t} \nonumber \\
    &\leq {\{ \frac{\gamma}{2\beta} a_{t-1} + \frac{\gamma}{2K} \}+a_t + \frac{\gamma}{K}} \nonumber 
    \end{align}

    We Observe $\phi(K) \leq \psi(K) \leq 3 \phi(K)$ and hence can rewrite our claim as : $\phi(K) \leq O(\frac{\log \log K}{\log K})$.
    
    We show an adversarial sequence to establish the upper bound. 

    Let the sequence $b_t$ be $\ell,\ell r,..,\ell {r^\tau},0,\ell,\ell r,..,\ell {r^\tau}$. We will calculate $\phi(K)$ based on a single period. Recall  $\ell = \frac{\gamma}{2K}$, $u=\frac{\gamma}{2} + \frac{\gamma}{2K}$.
    Note $l r^{\tau} = u$, so $\tau = \log_r (\frac{u}{l})$.
    \begin{eqnarray}
        \phi(K) \leq \frac{\tau}{\tau+1} \frac{1}{1+2r} + \frac{1}{\tau+1} \nonumber\\
        = \frac{\tau}{\tau+1} \frac{1}{1+2r} + \frac{1}{\tau+1} \left[ \frac{1}{1+2r} + \frac{2r}{1+2r} \right] \nonumber \\
        = \frac{1}{1+2r} + \frac{2r}{2r+1} \frac{1}{\tau+1} \nonumber \\
        = \frac{1}{1+2r} + \frac{2r}{2r+1} \cdot \frac{\log r}{\log(cr)} \nonumber
    \end{eqnarray}
    where $c = \frac{u}{l} = {K+1}$. Choosing $r = \log K$, we get
    \begin{eqnarray}
    \phi(K) \leq \frac{1}{1+2 \log K} + \frac{2 \log K}{2 \log K+1} \cdot \frac{\log \log K}{\log(K+1)} \nonumber \\
     = O \left( \frac{\log \log K}{\log  K} \right)
    \end{eqnarray}
\end{proof}

\subsection{Proof of Theorem~\ref{thm:tfpproj}}
\label{app:proof_trip}
\subsubsection{Strategy space inequality}
\label{sec:proof_ss1}
Let us define $f(\mathbf{\gamma})$ as follows:
\begin{align}
    f(\mathbf{\gamma}) = \min_{\mathbf{\beta}} \sum_{i=1}^K u_i \frac{\gamma_i}{\gamma_i+\beta_i}  \nonumber
\end{align}
We now show that $f(\mathbf{\gamma}) = (\sum_j \sqrt{u_j \gamma_j})^2$, using constrained Lagrange optimization.
\begin{align}
    f(\mathbf{\gamma}) &= \min_{\mathbf{\beta}} \sum_{i=1}^K u_i \frac{\gamma_i}{\gamma_i+\beta_i} \, s.t. \sum_{i=1}^K \beta_i = \beta \nonumber\\
    obj &= \sum_i u_i \frac{\gamma_i}{\gamma_i+\beta_i} + \lambda (\sum_{i=1}^K \beta_i - \beta) \nonumber\\
    \frac{\partial obj}{\partial \beta_i} &= \frac{-u_i\gamma_i}{(\gamma_i + \beta_i)^2} + \lambda = 0 \nonumber\\ 
    \gamma_i + \beta_i = \frac{\sqrt{u_i\gamma_i}}{\sqrt{\lambda}} \, &\text{and} \, \sqrt{\lambda} = \sum_{i=1}^{K}\sqrt{u_i\gamma_i} \nonumber 
\end{align}

substituting the value of $\lambda$ in $f(\mathbf{\gamma})$, we get:
\begin{align}
    f(\mathbf{\gamma}) &= \sum_{i=1}^K (\sqrt{u_i\gamma_i} \sum_{i=1}^K \sqrt{u_i\gamma_i}) = (\sum_{i=1}^K \sqrt{u_i\gamma_i})^2 \nonumber
\end{align}

Let $\boldsymbol{v}$ denote a sorted list of $\boldsymbol{u}$ sorted in a descending order and $pos(u_i)$ denote the position of $u_i$ in $\boldsymbol{v}$ with $pos(\max u_i)=1$.

We define $\tilde{u_i}$ as follows:
\[
    \tilde{u}_i= 
\begin{cases}
    u_i ,& \text{if } pos(u_i)\leq s\\
    0,              & \text{otherwise}
\end{cases}
\]

substituting $\gamma_i = \gamma \frac{\tilde{u}_i}{\sum_{i=1}^K \tilde{u}_i}$, we get:
\begin{align}
    f(\mathbf{\gamma}) = \frac{\gamma}{\sum_{i=1}^K \tilde{u}_i} (\sum_{i=1}^K \tilde{u}_i)^2 &= \gamma \sum_{i=1}^K \tilde{u}_i \geq \gamma \frac{s}{K} \sum_{i=1}^K u_i \nonumber\\
    \max_{\mathbf{\gamma}} \min_{\mathbf{\beta}} \sum_{i=1}^{K} u_i \frac{\gamma_i}{\gamma_i+\beta_i} &\geq f(\mathbf{\gamma}) \geq \gamma \frac{s}{K} \sum_{i=1}^{K} u_i \label{eqn:ssp}
\end{align}

\subsubsection{Modified strategy space inequality}
Let us now modify the policy $\boldsymbol{\gamma}$ to ensure that $\gamma_i \geq \frac{b}{s} if \tilde{\gamma}_i > 0$.
Let $\tilde{\gamma_i} = \gamma \frac{\tilde{u}_i}{\sum_{i=1}^K \tilde{u}_i}$, we have $ f(\boldsymbol{\tilde{\gamma}}) \geq \frac{s}{K}\gamma \sum_{i=1}^K u_i$
let us define $\boldsymbol{\gamma}$ as follows:
\begin{align}
    \boldsymbol{\gamma} = \frac{1}{1+q/\gamma} Proj(C_{q/s},\boldsymbol{\tilde{\gamma}}), \nonumber
\end{align}
where $Proj(C_{q/s}, \cdot)$ is a projection of the $s$ non-zero values of $\tilde{\gamma}_i$to the set $C_{q/s} = [q/s,1]^s$, the projection will cause at most $s$ values to grow by $\frac{q}{s}$. Thus, we normalize all values by $1+q/\gamma$.
We need to ensure that $\gamma_i \geq \frac{b}{s} \forall i \in [K]$, hence we get the inequality:
\begin{align}
    \frac{q/s}{1+q/\gamma} \geq \frac{q/s}{1+2q} &\geq \frac{b}{s} \Rightarrow q \geq \frac{b}{1-2b} \nonumber
\end{align}
we set $q = \frac{b}{1-2b}$, our modified $\boldsymbol{\gamma}$ now satisfies $\gamma_i \geq b/s$. 

We now have the following inequality
\begin{align}
    f(\boldsymbol{\gamma}) &\geq \frac{1}{1+q/\gamma}f(\boldsymbol{\tilde{\gamma}}) \geq (1-2b)f(\boldsymbol{\tilde{\gamma}}) \nonumber \\
    \min_{\boldsymbol{\beta}} \sum_i u_i \frac{\gamma_i}{\gamma_i+\beta_i}  &\geq \frac{s}{K} \gamma(1-2b)   \sum_{i=1}^K u_i\nonumber
\end{align}

\subsubsection{Stochastic strategy space inequality}

We define the event $E_i(t): \Gamma_i(t) \geq cN\gamma_i(t)$, and $\boldsymbol{E(t)}: \bigcap_{i=1}^K E_i(t)$

Let us compute the tail bound on $\Gamma_i(t)$ using the divergence bound on binomial distributions, we get
\begin{align}
    P(\Gamma_i(t)&\leq cN\gamma_i(t)) \leq e^{-nD(c\frac{\gamma_i(t)}{\gamma}||\frac{\gamma_i(t)}{\gamma})} \nonumber \\ D(c\frac{\gamma_i(t)}{\gamma}||\frac{\gamma_i(t)}{\gamma})
    &= c\frac{\gamma_i(t)}{\gamma} \log(c) + (1-c\frac{\gamma_i(t)}{\gamma}) \log(\frac{1-c\frac{\gamma_i(t)}{\gamma}}{1 - \frac{\gamma_i(t)}{\gamma}}) \nonumber \\
    &\geq (-c+c\log c+1)\frac{\gamma_i(t)}{\gamma} \nonumber\\
    &\geq (-c+c\log c+1)\gamma_i(t) \nonumber \\
    &\geq (-c+c\log c+1)\frac{b}{K} \nonumber\\
    P(\Gamma_i(t)&\leq cN\gamma_i(t)) \leq e^{-n\frac{b}{K}(-c+c\log c+1)}.  \label{eqn:inclusive2_divergence1}
\end{align}

We use the divergence bounds derived above to lower bound the event probability
\begin{align}
    P(\boldsymbol{E(t)}) &= 1-P(\boldsymbol{E(t)}^c) \geq 1 - \sum_{i=1}^K P(E_i(t)^c) \label{eqn:inclusive2_unionbound} \\
    &\geq 1 - s e^{-n\frac{b}{s}(-c + c\log c +1)} \label{eqn:inclusive2_divergence2}
\end{align}

where equation \ref{eqn:inclusive2_unionbound} uses a union bound amd the fact that $P(E_i(t)^c)=0$ for $K-s$ shards which have $\gamma_i=0$ and equation \ref{eqn:inclusive2_divergence2} is derived from equation \ref{eqn:inclusive2_divergence1}.

\begin{align}
    \mathbb{E}_{\boldsymbol{\Gamma(t)}}&\left[\min_{\boldsymbol{\beta(t)}} \sum_i u_i(t-1) \frac{\Gamma_i(t)}{\Gamma_i(t)+N\beta_i(t)}\right] \nonumber\\ 
    &\geq P(\boldsymbol{E(t)}) \min_{\boldsymbol{\beta(t)},\boldsymbol{E(t)}} \sum_i u_i(t-1) \frac{\Gamma_i(t)}{\Gamma_i(t)+N\beta_i(t)} \nonumber\\
    &\geq P(\boldsymbol{E(t)})(\sum_{i=1}^K \sqrt{u_ic\gamma_i})^2 \nonumber\\
    &\geq  P(\boldsymbol{E(t)})c(1-2b)\gamma\sum_{i=1}^K u_i \nonumber\\
    &\geq (1 - s e^{-n\frac{b}{s}(-c + c\log c +1)})c(1-2b) \gamma\sum_{i=1}^K \tilde{u}_i \nonumber\\
    &\geq (1 - s e^{-n\frac{b}{s}(-c + c\log c +1)})c(1-2b) \frac{s}{K}\gamma\sum_{i=1}^K u_i \nonumber \\
    \mathbb{E}_{\boldsymbol{\Gamma(t)}}&\left[\min_{\boldsymbol{\beta(t)}} \boldsymbol{u(t-1)}.\boldsymbol{r(t)} \right] \geq  h\sum_{i=1}^K u_i \nonumber
\end{align}
We want to lower bound $(1 - s e^{-n\frac{b}{s}(-c + c\log c +1)}) \geq 1-\epsilon$, thus we need
\begin{align}
    s e^{-n\frac{b}{s}(-c + c\log c +1)} &\leq \epsilon \Rightarrow
    n \geq \frac{s\log(\frac{s}{\epsilon})}{b(-c + c\log c +1)} \nonumber \\
    N &\geq \frac{s\log(\frac{s}{\epsilon})}{\gamma b(-c + c\log c +1)} \label{eqn:klogk}
\end{align}

\subsubsection{Approach to the convex set}
We will now show that $\boldsymbol{\bar{r}(t)}$ approaches $C_{h}$ with $h=(1 - s e^{-n\frac{b}{s}(-c+clogc+1)})c(1-2b)\frac{s}{K}\gamma$. 

Let $\pi_{C_{h}}(t)$ be the projection of $\boldsymbol{\bar{r}(t)}$ on $C_{h}$. Let us define a halfspace $\boldsymbol{H}_{t+1}$ formed by the hyperplane $\boldsymbol{P}_{t+1}$, such that $\boldsymbol{P}_{t+1}$ is normal to $(\boldsymbol{\bar{r}(t)}-\boldsymbol{\pi_{C_h}(t)})$ and passes through $\pi_{C_{h}}$ and $\boldsymbol{H}_{t+1}$ contains $C_{h}$. The variables defined above satisfy the following equations: 
\begin{align}
    \boldsymbol{\pi_{C_h}(t)} &= arg \min_{\boldsymbol{y}\in C_h} || \boldsymbol{y}-\boldsymbol{\bar{r}}(t)|| \nonumber\\
    \pi_{C_{h}}(t)_i &= h I_{\{ \bar{r_i}(t)<h \}} + \bar{r_i}(t)I_{\{ \bar{r_i}(t)<h \}} \nonumber\\
    \boldsymbol{\pi_{C_h}(t)}-\mathbf{\bar{r}(t)} &= \boldsymbol{u(t)} \quad u_i(t) = (h-\bar{r}_i(t)) \nonumber \\
    \boldsymbol{P}_{t+1}(\boldsymbol{x}) &: \sum_{i=1}^{K}(h-\bar{r_i}(t))_+ x_i - h \sum_{i=1}^{K}(h-\bar{r_i}(t))_+ = 0 \nonumber\\
    &\mbox{such that }\quad  \boldsymbol{u\cdot x} - h\sum_{i=1}^K u_i = 0 \nonumber\\
    \boldsymbol{H}_{t+1}(\boldsymbol{x}) &: \sum_{i=1}^{K}(h-\bar{r_i}(t))_+ x_i - h \sum_{i=1}^{K}(h-\bar{r_i}(t))_+ \geq 0 \nonumber\\
    &\mbox{such that }\quad \boldsymbol{u\cdot x} - h\sum_{i=1}^K u_i \geq 0 \nonumber
\end{align}

Let us define $d_t$ as the distance of $\boldsymbol{\bar{r}}(t)$ from the convex set $C_h$, i.e. $d_t = || \boldsymbol{\bar{r}}(t) - \boldsymbol{\pi}_{C_h}(t) ||$ , $d(\boldsymbol{a,b}) = ||\boldsymbol{a} - \boldsymbol{b}||$ for any $\boldsymbol{a},\boldsymbol{b} \in \mathbf{R}^K$. 

\begin{align}
    d^{2}_{t+1} &= d^2(\mathbf{\bar{r}(t+1)},\boldsymbol{\pi_{C_h}(t+1)}) \leq d^2(\mathbf{\bar{r}(t+1)},\boldsymbol{\pi_{C_h}(t)}) \nonumber\\
    &= \left\lVert \mathbf{\bar{r}(t+1)} - \boldsymbol{\pi_{C_h}(t)} \right\rVert^2_2 \nonumber\\
    &= \left\lVert \frac{t}{t+1}\mathbf{\bar{r}(t)} + \frac{1}{t+1} \mathbf{r(t+1)} - \boldsymbol{\pi_{C_h}(t)} \right\rVert^2_2 \nonumber\\
    &= \left\lVert \frac{t}{t+1}(\mathbf{\bar{r}(t)} -\boldsymbol{\pi_{C_h}(t)}) + \frac{1}{t+1}( \mathbf{r(t+1)} - \boldsymbol{\pi_{C_h}(t)}) \right\rVert^2_2 \nonumber\\
    &= (\frac{t}{t+1})^2 \left\lVert  \mathbf{\bar{r}(t)} -\boldsymbol{\pi_{C_h}(t)} \right\rVert^2_2 + (\frac{1}{t+1})^2 \left\lVert  \mathbf{r(t+1)} - \boldsymbol{\pi_{C_h}(t)} \right\rVert^2_2 \nonumber\\ 
    &+ \frac{2t}{(t+1)^2}(\mathbf{\bar{r}(t)} -\boldsymbol{\pi_{C_h}(t)}).(\mathbf{r(t+1)} - \boldsymbol{\pi_{C_h}(t)}) \nonumber 
\end{align}

\begin{align}
    (t+1)^2 d^2_{t+1} - t_2 d^2_t  &\leq \left\lVert  \mathbf{r(t+1)} - \boldsymbol{\pi_{C_h}(t)} \right\rVert^2_2 \nonumber \\
    &+ 2t*((\boldsymbol{\pi_{C_h}(t)}-\mathbf{\bar{r}(t)} ).( \boldsymbol{\pi_{C_h}(t)}-\mathbf{r(t+1)}))\nonumber
\end{align}

we know the following equations:
\begin{align}
    \left\lVert  \mathbf{r(t+1)} - \boldsymbol{\pi_{C_h}(t)} \right\rVert^2_2 &\leq h^2K \nonumber\\
    (\boldsymbol{\pi_{C_h}(t)}-\mathbf{\bar{r}(t)}).\boldsymbol{\pi_{C_h}(t)}) &= h\sum_{i=1}^K u_i \nonumber\\
    (\boldsymbol{\pi_{C_h}(t)}-\mathbf{\bar{r}(t)}).\boldsymbol{\pi_{C_h}(t)}) &\leq \mathbb{E}_{\Gamma_i(t+1)}\left[\sum_i u_i(t) r_i(t+1)\right] \nonumber 
\end{align}

Summing terms for $t\in[T]$, we get
\begin{align}
    d^2_{T} &\leq h^2\frac{K}{T} + \frac{2}{T}\sum_{t=1}^{T-1}\frac{t}{T}(\mathbb{E}_{\Gamma_i(t)}\left[\sum_i u_i(t-1) r_i(t)\right] - (\boldsymbol{u(t-1)}).\mathbf{r(t)}) \nonumber
\end{align}

The term $Y_t = (\mathbb{E}_{\Gamma_i(t)}\left[\sum_i u_i(t-1) r_i(t)\right] - ((\boldsymbol{u(t-1)}).\mathbf{r(t)})$ is a martingale difference sequence w.r.t. history at time $t$ and $|Y_t|\leq 2 h s$

Given $\epsilon_m>0$, by the Azuma- Hoeffding inequality we have:
\begin{align}
    \mathbb{P}\left( \frac{1}{T} \left\lVert \sum_{t=1}^{T-1} Y_t \right\rVert > \epsilon_m \right) \leq 2 e^{ -\frac{T\epsilon_m^2}{8h^2s^2}} \nonumber
\end{align}

Let us set $\epsilon_m = 2hs\sqrt{\frac{2}{T}\log(\frac{2}{\delta})}$, we get
\begin{align}
    \mathbb{P}\left( \frac{1}{T} \left\lVert \sum_{t=1}^{T-1} Y_t \right\rVert > \epsilon_m \right) \leq \delta \nonumber
\end{align}

Thus with a high probability of $1-\delta$, we have distance to the convex set reducing as:
\begin{align}
     d^2_{T} &\leq h^2\frac{K}{T} + 4hs\sqrt{\frac{2}{T}\log(\frac{2}{\delta})} \nonumber
\end{align}

\subsection{Proof of Proposition~\ref{prop:advlatency}}
\label{app:latencyboundedadversary}

At $t=1$ the adversary is focused on $\frac{K}{\log K}$ shards, let those shards form a set $S_1$. Let the honest node allocate themselves as $\boldsymbol{\gamma(2)}$
\begin{align}
    \frac{1}{|S_1|}\sum_{i\in S_1} \gamma_i(2) \leq \frac{1-\beta}{|S_1|} \nonumber
\end{align}
The median of ${\gamma_i(2)}$ is less than twice its mean, thus
$\text{median} \leq \frac{1(1-\beta)(\log K)}{K}$. This implies that the last $\frac{K}{2\log K}$ shards (arranged in descending order of their honest fraction) will have honest fraction less than $\frac{1(1-\beta)(\log K)}{K}$.

The adversary now spreads only to $\frac{K}{(\log K)^2}$ lowest performing shards, this set is made up of the subset of the $\frac{K}{2\log K}$ shards discussed above since $\frac{K}{2\log K}\leq \frac{K}{(\log K)^2}$ (for $K>e^2$). 

We can again show that the adversary allocates as $\boldsymbol{\gamma(3)}$ and show that the last $\frac{K}{2^2\log K}$ shards (arranged in descending order of their honest fraction) will have honest fraction less than $\frac{1(1-\beta)(\log K)}{K}$ and the adversary now spreads only to $\frac{K}{(\log K)^3}$ lowest performing shards, this set is made up of the subset of the $\frac{K}{2^2\log K}$ shards discussed above since $\frac{K}{2\log K}\leq \frac{K}{(\log K)^2}$ (for $K>e^2$).

The attack continues till $\frac{K}{(\log K)^{\tau}}=1$, the adversary cannot concentrate further. Solving the above equation, we get $\tau=\frac{\log K}{\log \log K}$, completing the proof. 

\section{Other Preliminaries}
\subsection{Verifiable Random Function \label{app:vrf}}

Verifiable Random Functions (VRF), first introduced in \cite{vrf}, generates a pseudorandom number with a proof of its correctness. A node with a secret key $sk$ can call {\sc VRFprove}$(\cdot,sk)$ to generates a pseudorandom {\em output} $F_{sk}(\cdot)$ along with a {\em proof} $\pi_{sk}(\cdot)$. Other nodes that have the proof and the corresponding public key $pk$ can check that the output has been generated by VRF, by calling {\sc VRFverify}$(\cdot,{\rm output},\pi_{sk}(\cdot),pk)$. An efficient implementation of VRF was introduced in \cite{vrf2}. This ensures that the output of a VRF is computationally indistinguishable from a random number even if the public key $pk$ and the function {\sc VRFprove} is revealed.

%%% Local Variables:
%%% mode: latex
%%% TeX-master: "main"
%%% End:

\end{document}